\documentclass[%
reprint,
aps,
superscriptaddress,
amsmath,amssymb,
]{revtex4-2}

\usepackage{graphicx}
\usepackage{dcolumn}
\usepackage{bm}
\usepackage[dvipsnames]{xcolor}
\usepackage{changes}
\usepackage{xr-hyper}
\usepackage{hyperref}

\makeatletter

\def\p@subsection{}
\def\@sec@style{\bfseries}  
\renewcommand{\fnum@figure}{\textbf{Figure \thefigure}}
\renewcommand{\fnum@table}{\textbf{Table \thetable}}
\renewcommand{\thetable}{\arabic{table}}
\makeatother

\begin{document}
	\preprint{APS/123-QED}
	
	\title{The influence of multi-dimensionality and off-diagonal non-Markovian friction coupling on coarse-grained dynamics}
	
	\author{Henrik Kiefer}
	\author{Cihan Ayaz}
	\author{Benjamin A. Dalton}
	\author{Roland R. Netz}
	\altaffiliation{Author to whom any correspondence should be addressed: rnetz@physik.fu-berlin.de}%
	\affiliation{%
		Department of Physics, Freie  Universität  Berlin,  Arnimallee  14,  14195  Berlin,  Germany.
	}%
	
	\date{\today}

	\begin{abstract}
		Coarse-graining complex molecular systems to lower-dimensional reaction coordinates is a powerful approach for capturing their effective dynamics. The generalized Langevin equation (GLE) provides an exact framework for modeling coarse-grained dynamics, and is particularly useful when non-Markovian effects are significant. While one-dimensional GLE models are commonly used, many systems require multi-dimensional reaction coordinates to account for coupled dynamics. 
		Here, we study the GLE formalism for multi-dimensional reaction coordinates, incorporating a memory matrix to quantify  non-Markovian frictional coupling between coordinates, and a multi-dimensional potential. Using the GLE model, in conjunction with a multi-dimensional Markovian embedding scheme, we investigate different systems that are characterized by two-dimensional reaction coordinates, namely the dihedral dynamics of pentane and alanine dipeptide, obtained from molecular dynamics simulations in explicit water. We identify significant off-diagonal friction couplings arising from intramolecular and hydrodynamic interactions. Unlike previous studies, our results highlight the critical role of different terms in the multi-dimensional GLE in accurately capturing key dynamical properties, including mean first-passage times and mean-squared displacements, particularly in systems with coupled non-Markovian coordinates. 
		
	\end{abstract}
	
	\maketitle
	
	\newpage
	
	\section{Introduction}
	The modeling of interacting many-particle systems necessitates calculating interactions and tracking the temporal evolution of numerous degrees of freedom \cite{van1998remarks, huang2009introduction}. To effectively capture and interpret the system's dynamics, low-dimensional reaction coordinates are often employed \cite{kardar2007statistical, balian2007microphysics}. This coarse-graining approach eliminates the need to solve equations of motion for all degrees of freedom, while, depending on the coarse-graining method, may still provide an accurate description of the system's effective kinetics based on the reaction coordinates. The utility of this methodology has been demonstrated across various fields, including polymer dynamics \cite{hijon2010mori, karimi2012good}, chemical reaction modeling \cite{adelman1980generalized, straub1987calculation,canales1998generalized, bagchi1983effect}, and conformational dynamics, such as protein folding \cite{plotkin1998non, satija2019generalized, lange2006collective}.\\
	\indent Reducing the degrees of freedom to scalar observables introduces non-Markovianity, commonly referred to as memory effects, into the coarse-grained dynamics \cite{zwanzig1961memory, mori1965transport, carof2014two}. The generalized Langevin equation (GLE) has become a widely adopted framework for constructing robust molecular models that incorporate these memory effects.
	By accounting for time-dependent friction, GLE models offer more accurate kinetic predictions compared to standard Langevin models \cite{bagchi1983effect,canales1998generalized,satija2019generalized,chorin2002optimal,ayaz2021non,klippenstein2021introducing,brunig2022timedependent,brunig2022pair,dalton2022protein, dalton2024role, dalton2024memory, kiefer2024predictability}.
	\\ \indent The GLE is often applied to one-dimensional (1D) reaction coordinates derived from particle distances or the center of mass of atom groups \cite{berne1990dynamic,lange2006collective, ceriotti2010colored,ayaz2022generalized,vroylandt2022gle}. Projecting all degrees of freedom onto a single relevant observable enables accurate modeling and kinetic predictions for transition processes like protein folding \cite{ayaz2021non, dalton2022protein, dalton2024role} or molecular vibrational spectra \cite{brunig2022timedependent}, only when memory effects are appropriately included. Nevertheless, identifying optimal 1D reaction coordinates remains an active area of research \cite{best2013native, berezhkovskii2018single} and is not always straightforward. Challenges include coupling of rotational and translational degrees of freedom or correlations between subdomain rotations and cis-trans isomerization reactions in molecules \cite{abrash1990viscosity,bagchi2012molecular}. For isomerization, the evolution of a reaction coordinate on a potential landscape may be influenced by interactions with other reactive or non-reactive coordinates \cite{langer1969statistical, grote1981reactive, van1981stochastic,acharya2023diffusion}, and can for 1D GLE formulations lead to slowly decaying memory functions and non-Gaussian stochastic forces, that are difficult to deal with in practice \cite{mazur1991and, vroylandt2022position}.
	\\ \indent Multi-dimensional transition dynamics at a Markovian level has been extensively studied \cite{van1982reactive,bryngelson1989intermediates,langer2000theory}, but interactions with the environment can introduce additional complexity through non-Markovian friction \cite{kraft2013brownian, acharya2021rate,acharya2022non}. The non-Markovian behavior of particles, which we aim to capture with suitable reaction coordinates, can be further influenced by friction-force coupling via hydrodynamic interactions \cite{deutch1971molecular, ermak1978brownian, van1981stochastic, reichert2004hydrodynamic, jung2017frequency} or by intramolecular degrees of freedom \cite{ansari1992role, de2014molecular, daldrop2018butane}, which create correlations between forces acting on reaction coordinates, potentially accelerating or slowing down transition dynamics \cite{van1982reactive,acharya2022non}. \\
	\indent Numerous studies have explored systems with friction coupling between degrees of freedom within a multi-dimensional Markovian Langevin equation to describe the conformational dynamics of small peptides and proteins \cite{Hegger09, Schaudinnus15, Lickert20}. However, these approaches usually rely on memoryless coordinates, operating under the assumption of time-scale separation, wherein the coarse-grained coordinates relax much slower than the environment. 
	In contrast, investigating the non-Markovian dynamic interplay between inter- or intramolecular degrees of freedom is essential for accurately modeling complex system dynamics also within a multi-dimensional non-Markovian GLE framework, which includes friction matrices with off-diagonal terms and correlations between stochastic force components \cite{zwanzig2001nonequilibrium,lee2019multi,she2023data,lyu2023construction}.
	\\ \indent Here, we use a multi-dimensional GLE approach to general coarse-grained reaction coordinates that are non-linear functions of microscopic coordinates. Instead of employing independent, uncoupled GLE models for each coordinate \cite{xie2024coarse, xie2024ab}, the GLE model incorporates a memory kernel matrix with non-zero off-diagonal elements, capturing correlations between frictional forces acting on different reaction coordinates. This formalism includes a mean-force term derived from a multi-dimensional non-linear potential landscape, obtained by integrating out all degrees of freedom except for the relevant observables.\\
	\indent We first present a method to compute the entries of the memory matrix from reaction coordinate trajectories. Subsequently, we introduce a GLE simulation technique based on a Markovian embedding scheme that enables efficient numerical solutions of the multi-dimensional GLE once its parameters have been determined. 
	\\
	\indent We apply the extraction algorithm to time-series data derived from water-explicit molecular dynamics (MD) simulations on dihedral angle dynamics. First, we examine the dihedral angle dynamics of a pentane molecule, which involves two dihedral angles and presents a two-dimensional (2D) problem.
	\\
	\indent By
	simulating the GLE using a 2D Markovian embedding with parameters from the extracted memory kernel matrix results in a much better prediction of MD mean first-passage times than 1D GLE simulations. We analyze in detail how off-diagonal entries in the memory kernel matrix and the 2D potential landscape affect the dynamics. As a second application, we study the 2D dihedral dynamics of an alanine dipeptide molecule in water. 
	\\\indent We extend the work in Ref.~\cite{lee2019multi}, where it was shown that a 2D GLE is accurate for the dihedral dynamics of alanine dipeptide, by comparing the accuracy of 2D GLEs to uncoupled 1D GLE formulations, which highlights the importance of off-diagonal friction and multi-dimensional confinement. To simulate the GLE, we map the GLE onto a high-dimensional Markovian embedding while assuming the memory kernel to be a sum of matrix exponentials. This mapping procedure is a simplified version of the methodology proposed in Ref.~\cite{lee2019multi}, which captures the temporal decay in the memory kernel and preserves the coarse-grained system dynamics.
	\section{Theory and methods}
	\subsection{The multi-dimensional GLE and parameter extraction}
	\label{sec:extraction}
	We consider the phase space $\Omega$ of a system of $N$ interacting atoms or particles in three-dimensional space. The time evolution of the microstate $\vec{\omega}(t)$, a $6N$-dimensional vector of positions $\vec{q}_i$ and momenta $\vec{p}_i$ for $i = 1, 2, ..., N$, is governed by the Hamilton equations of motion
	\begin{equation}
		\label{eq:phase_space}
		\Dot{\vec{\omega}}(t) = \mathcal{L} \vec{\omega}(t),
	\end{equation}
	where $\mathcal{L}$ is the Liouville operator, and the initial state is $\vec{\omega}(0) = \vec{\omega}_0$.
	\\ \indent We coarse-grain the dynamics in equation~\eqref{eq:phase_space} by choosing $n < N$ observables, represented by a vector $\vec{x}$. The coarse-graining approach follows the projection operator formalism of R. Zwanzig \cite{zwanzig1961memory} and H. Mori \cite{mori1965transport}. We adopt the recent derivation of the GLE using hybrid projection \cite{ayaz2022generalized, vroylandt2022gle}. For a set of coarse-grained observables $\Vec{x}(t) = \bigl(x_1(t), x_2(t), \dots, x_n(t)\bigr)^T$ and their velocities $\Dot{\Vec{x}}(t)$, the multi-dimensional GLE reads \cite{ayaz2022generalized,vroylandt2022gle}
	\begin{equation}
		\label{eq:multi_GLE_hybrid}
		\Ddot{\Vec{x}}(t) = -\Vec{F}_{\text{eff}}\bigl(\Vec{x}(t)\bigr) - \int_0^t \:ds\:\Hat{\Gamma}^*(s) \Dot{\Vec{x}}(t-s) + \Vec{F}^*_\text{R}(t),
	\end{equation}
	where 
	\begin{align}
		\label{eq:force_effective}
	\Vec{F}_{\text{eff}}\bigl(\Vec{x}(t)\bigr) = & \: k_B T \Hat{M}^{-1}\bigl(\Vec{x}(t)\bigr) \vec{\nabla} \ln\Hat{M}\bigl(\Vec{x}(t)\bigr)  \\ \nonumber
		&+ \Hat{M}^{-1}\bigl(\Vec{x}(t)\bigr) \vec{\nabla} U\bigl(\Vec{x}(t)\bigr)
	\end{align}
	is the conservative mean force acting on $\vec{x}$. $\Hat{M}(\vec{x})$ is the mass matrix, $U(\vec{x})$ is the potential of mean force (PMF), $k_BT$ is the product of Boltzmann constant and temperature $T$, $\Hat{\Gamma}^*(t)$ is the memory kernel matrix, and $\Vec{F}^*_\text{R}(t)$ is the random force vector. The effect of the irrelevant coordinates is captured entirely by $\Vec{F}^*_\text{R}(t)$. For an equilibrium system, the memory kernel governs $\Vec{F}^*_\text{R}(t)$ via the approximate relation \cite{ayaz2022generalized, vroylandt2022gle}
	\begin{equation}
		\label{eq:fdt_multi_hybrid}
		\langle \Vec{F}^*_\text{R}(t)(\Vec{F}_\text{R}^*)^T(0) \rangle = \langle \Dot{\Vec{x}}\Dot{\Vec{x}}^T \rangle\Hat{\Gamma}^*(t),
	\end{equation}
	where $\langle ... \rangle$ denotes an ensemble average.
	\noindent The inverse mass matrix $\Hat{M}^{-1}({\Vec{x}})$ is determined by the conditional average \cite{darve2001calculating, lee2019multi, ayaz2022generalized}
	\begin{equation}
		\label{eq:mass_equip_pos_dep}
		\Hat{M}^{-1}(\Vec{x}) = (k_BT)^{-1} \langle \Dot{\Vec{x}}\Dot{\Vec{x}}^T \rangle_{\vec{x}}.
	\end{equation}
	We will assume the entries of the mass matrix $\Hat{M}$ to be constant and to not depend on the reaction coordinates, which is an approximation we will test using our MD data. In this case, the effective mass matrix is given by an unconstrained average
	\begin{equation}
		\label{eq:mass_equip}
		\Hat{M}^{-1} = (k_BT)^{-1} \langle \Dot{\Vec{x}}\Dot{\Vec{x}}^T \rangle,
	\end{equation}
	and the force in equation~\eqref{eq:force_effective} only includes the potential gradient term $\Vec{\nabla}U(\vec{x})$. The GLE in equation~\eqref{eq:multi_GLE_hybrid} reduces to, by multiplying both sides with $\Hat{M}$,
	\begin{equation}
		\label{eq:multi_GLE}
		\Hat{M} \Ddot{\Vec{x}}(t) = - \vec{\nabla}U\bigl(\Vec{x}(t)\bigr) - \int_0^t ds\:\Hat{\Gamma}(s) \Dot{\Vec{x}}(t-s)  + \Vec{F}_\text{R}(t),
	\end{equation}
	where $\Hat{\Gamma}(t) = \hat{M}\hat{\Gamma}^*(t)$ and $\Vec{F}_\text{R}(t) = \hat{M}\Vec{F}^*_\text{R}(t)$. The approximate relation in equation~\eqref{eq:fdt_multi_hybrid} is modified as
	\begin{equation}
		\label{eq:fdt_multi}
		\langle \Vec{F}_\text{R}(t)\Vec{F}_\text{R}^T(0) \rangle = k_BT\:\Hat{\Gamma}(t).
	\end{equation}
	For vanishing off-diagonal entries in the memory kernel matrix $\Hat{\Gamma}(t)$ and the effective mass matrix $\Hat{M}$, and a PMF with no coupling terms, which reads
	\begin{equation}
		\label{eq:additive_pmf}
		U(\Vec{x}) = U_1(x_1) + U_2(x_2) + \dots + U_n(x_n),
	\end{equation}
	the GLE in equation~\eqref{eq:multi_GLE} separates into uncoupled one-dimensional equations, as we will discuss later.\\
	\indent Similar to the one-dimensional case \cite{daldrop2018butane, ayaz2021non, berne1970calculation}, we use an iterative formula for extracting the memory kernel matrix from discrete time-series data $\vec{x}(t)$ \cite{lee2019multi}. Previous work in the one-dimensional case has shown that extracting the memory kernel's running integral, $\hat{G}(t) = \int_0^t ds\: \hat{\Gamma}(s)$, from a Volterra equation produces more stable results \cite{kowalik2019memory, ayaz2021non, brunig2022timedependent}. We here generalize the $\hat{G}$-approach to the multi-dimensional case. The iterative equation for the discretized running integral over the memory kernel matrix $\Hat{G}_i = \hat{G}(i\Delta t)$, with $\Delta t$ being the time resolution of $\vec{x}(t)$, is given by
	\begin{align}
		\label{eq:G-iter}
		\Hat{G}_i  =& \Bigl\lbrack C_i^{\Vec{\nabla}U\Vec{x}} - C_0^{\Vec{\nabla}U\Vec{x}} - \Hat{M}( C_i^{\Dot{\Vec{x}}\Dot{\Vec{x}}} - C_0^{\Dot{\Vec{x}}\Dot{\Vec{x}}}) \\\nonumber &- \Delta t \sum_{j=1}^{i-1}\Hat{G}_j C_{i-j}^{\Dot{\Vec{x}}\Dot{\Vec{x}}}\Bigr\rbrack \cdot (\frac{1}{2}\Delta t\:C_0^{\Dot{\Vec{x}}\Dot{\Vec{x}}})^{-1},
	\end{align}
	with $\Hat{G}_0 = 0$, and $C_i^{\Vec{\nabla}U\Vec{x}} = \langle \Vec{\nabla}U\bigl(\Vec{x}(i \Delta t)\bigr) \Vec{x}^T(0) \rangle$ and $C_i^{\Dot{\Vec{x}}\Dot{\Vec{x}}} = \langle \Dot{\Vec{x}}(i \Delta t) \Dot{\Vec{x}}^T(0)\rangle$ are correlation matrices. The derivation of the iteration scheme in equation~\eqref{eq:G-iter} is given in appendix \ref{app:iter}. From a time-series trajectory $\Vec{x}(t)$, we can extract the running integral $\Hat{G}(t)$ for a given time resolution $\Delta t$ and, by taking a derivative, obtain the memory kernel matrix $\Hat{\Gamma}(t)$. In appendix \ref{app:test_extraction}, we apply the extraction technique on a simple model system, demonstrating its robustness.\\
	\indent In equilibrium, the PMF $U(\Vec{x})$ is determined by the probability distribution $\rho(\Vec{x})$ as $U(\Vec{x}) = -k_BT \ln\rho(\Vec{x})$, in conjunction with cubic-spline interpolation \cite{Margolis2019}. 
	For some systems, the PMF decouples as in equation~\eqref{eq:additive_pmf}. However, this depends on the system and the choice of coordinates. Generally, it is necessary to compute the full multi-dimensional potential landscape $U(\Vec{x}) = U(x_1, x_2, \dots, x_n)$.
	
	\subsection{Multi-dimensional Markovian embedding}
	\label{sec:mapping}
	
	Equation~\eqref{eq:fdt_multi} complicates modeling of the random force vector $\Vec{F}_\text{R}(t)$, as its entries are correlated if the memory kernel matrix $\hat{\Gamma}(t)$ has non-zero off-diagonal entries. Following the approach in the one-dimensional case \cite{ceriotti2010colored, li2017computing, ayaz2021non, kappler2019non, lavacchi2020barrier}, we map the GLE in equation~\eqref{eq:multi_GLE} onto an embedding Markovian system in multiple dimensions, assuming a multi-exponential Ansatz for the memory kernel matrix, according to
	\begin{figure*}
		\centering
		\includegraphics[width=1
		\linewidth]{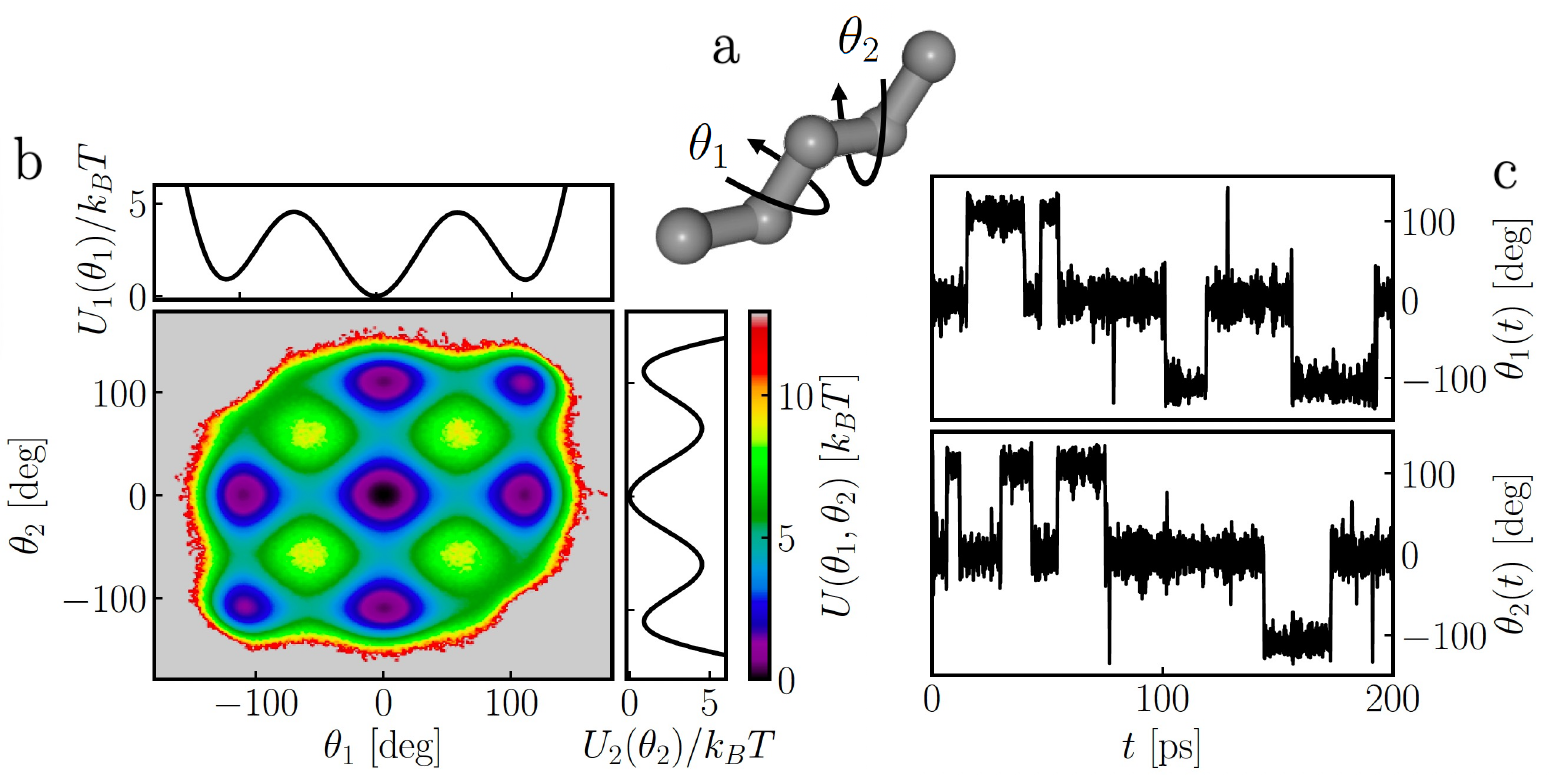}
		\caption{Two-dimensional dihedral observable of pentane. (a) Illustration of the two dihedral angles $\theta_1$ and $\theta_2$ in a pentane molecule. (b) 2D potential $U(\theta_1,\theta_2) = -k_BT\ln \rho(\theta_1,\theta_2)$ of the dihedral angles $\theta_1$ and $\theta_2$ from a solvated pentane MD simulation, together with the 1D landscapes $U_1(\theta_1)=-k_BT\ln \rho(\theta_1)$ and $U_2(\theta_2)=-k_BT\ln \rho(\theta_2)$. (c) Sample trajectories of both dihedral angles.}
		\label{fig:pentane_dihedrals_scheme}
	\end{figure*}
	\begin{align}
		\label{eq:Markovian2}
		\Hat{M} \Ddot{\Vec{x}}(t) &= - \Vec{\nabla}U\bigl(\Vec{x}(t)\bigr) -\sum_{i=1}^m \Hat{\gamma}_{i}\Hat{\tau}_{i}^{-1}\bigl(\Vec{x}(t)- \Vec{y}_i(t)\bigr) ,\\ \nonumber
		\Dot{\Vec{y}}_i(t) &= - \Hat{\tau}_{i}^{-1} \bigl(\Vec{y}_i(t) - \Vec{x}(t)\bigr) + \Vec{\eta}_i(t),
	\end{align}
	\noindent where $\Hat{\gamma}_{i}$ are the friction coefficient matrices and $\Hat{\tau}_{i}$ the memory time matrices. The total friction coefficient matrix $\hat{\gamma} = \sum_{i=1}^m \hat{\gamma}_i$ is thereby defined as the long-time limit of $\hat{G}(t)$, i.e. $\hat{\gamma} = \int_0^\infty ds\:\hat{\Gamma}(s)$. We assume that all matrices are invertible. In this Markovian embedding, the coupling of $\vec{x}$ to the auxiliary variables $\vec{y}_i$ generates the non-Markovian dynamics of $\vec{x}$, given that equation~\eqref{eq:fdt_multi} is fulfilled. The dynamics of $\vec{y}_i$ is overdamped; equations~\eqref{eq:Markovian2} are, therefore, a simplified version of the Markovian embedding proposed in Ref.~\cite{lee2019multi}. The random variables $\vec{\eta}_i(t)$ follow delta-correlated Gaussian processes with $\langle \vec{\eta}_i(t) \rangle = 0$ and second moment
	\begin{equation}
		\label{eq:Markovian3}
		\langle \Vec{\eta}_i(t) \Vec{\eta}_j^T(t') \rangle =  2k_BT\Hat{\gamma}_{i}^{-1} \delta_{ij} \delta(|t-t'|).
	\end{equation}
	Equations~\eqref{eq:Markovian2} result in the memory kernel matrix
	\begin{align}
		\label{eq:memory_matrix}
		\Hat{\Gamma}(t) = \sum_{i=1}^m \Hat{\gamma}_{i} \Hat{\tau}_{i}^{-1} e^{-t\Hat{\tau}_{i}^{-1}} = \sum_{i=1}^m \Hat{\Gamma}_{i}(t).
	\end{align}
	The derivation of equation~\eqref{eq:memory_matrix} is shown in appendix \ref{app:mapping}.
	\\
	\indent Once the friction coefficient matrices $\Hat{\gamma}_{i}$ and memory time matrices $ \Hat{\tau}_{i}$ are given, we can solve equations~\eqref{eq:Markovian2} numerically for an effective GLE simulation of the underlying dynamical problem.
	\subsection{Reduction to one-dimensional GLE models}
	\label{sec:mapping_1D}
	For uncoupled reaction coordinates, i.e. for a decoupled potential according to equation~\eqref{eq:additive_pmf} and diagonal $\Hat{\Gamma}$ and $\Hat{M}$ matrices, the GLE in equation~\eqref{eq:multi_GLE} reduces to decoupled 1D GLEs for the coordinates $x_k$ of $\vec{x}$, given by
	\begin{equation}
		\label{eq:1D_GLE}
		M_{kk} \Ddot{x}_k(t) = -   \nabla U_k\bigl(x_k(t)\bigr) - \int_0^t ds\:\Gamma_{kk}(s) \Dot{x}_k(t-s)  + F_\text{R}^k(t),
	\end{equation}
	where $\langle F_\text{R}^k(t)F_\text{R}^k(0)\rangle = k_BT \:\Gamma_{kk}(t)$, with $\Gamma_{kk}$ and $M_{kk}$ being the diagonal entries of $\hat{\Gamma}$ and $\hat{M}$, respectively. The iterative formula for the memory kernel in equation~\eqref{eq:G-iter} reduces to the one-dimensional form derived in Ref.~\cite{ayaz2021non}. If we parameterize the extracted memory kernel $\Gamma_{kk}(t)$ as a sum of $m$ exponential functions with friction coefficients $\gamma^k_i$ and memory times $\tau^k_i$
	\begin{eqnarray}
		\label{eq:fit_kernel}
		\Gamma_{kk}(t) = \sum_{i=1}^{m} \frac{\gamma^{k}_i}{\tau^k_i} e^{-t/\tau^k_{i}},
	\end{eqnarray}
	the GLE in equation~\eqref{eq:1D_GLE} is equivalent to the following Markovian embedding \cite{ayaz2021non, kappler2019non, lavacchi2020barrier}:
	\begin{align}
		\label{eq:markov_embedding_1D}
		M_{kk} \Ddot{x}_k(t) &= - \nabla U_k\bigl(x_k(t)\bigr) - \sum_{i=1}^{m} \frac{\gamma^k_i}{\tau^k_i}\bigl(x_k(t) - y^k_i(t)\bigr)  , \\ \nonumber
		\Dot{y}^k_i(t) &= - \bigl(y^k_i(t) - x_k(t)\bigr)/\tau^k_{i} + \eta^k_i(t).
	\end{align}
	The random forces $\eta^k_i(t)$ follow stationary Gaussian processes with zero mean and $\langle \eta^k_i(t) \eta^k_j(t')\rangle = 2 k_BT (\gamma^k_i)^{-1} \delta_{ij} \delta(|t-t'|)$. Using values of $\gamma^k_i$ and $\tau^k_i$, we can numerically solve the system of Markovian Langevin equations in equations~\eqref{eq:markov_embedding_1D} with a time step $\Delta t$ to obtain trajectories of $x_k$. This 1D embedding will be used to compare with our multi-dimensional GLE (equations~\eqref{eq:Markovian2}), testing the significance of off-diagonal friction coupling and coupled potentials that cannot be written in the form of equation~\eqref{eq:additive_pmf}.
	
	\section{Application of the multi-dimensional GLE to the dihedral angle dynamics in pentane}
	\label{sec:pentane_free}
	
	\begin{figure*} 
		\centering
		\includegraphics[width=0.7
		\linewidth]{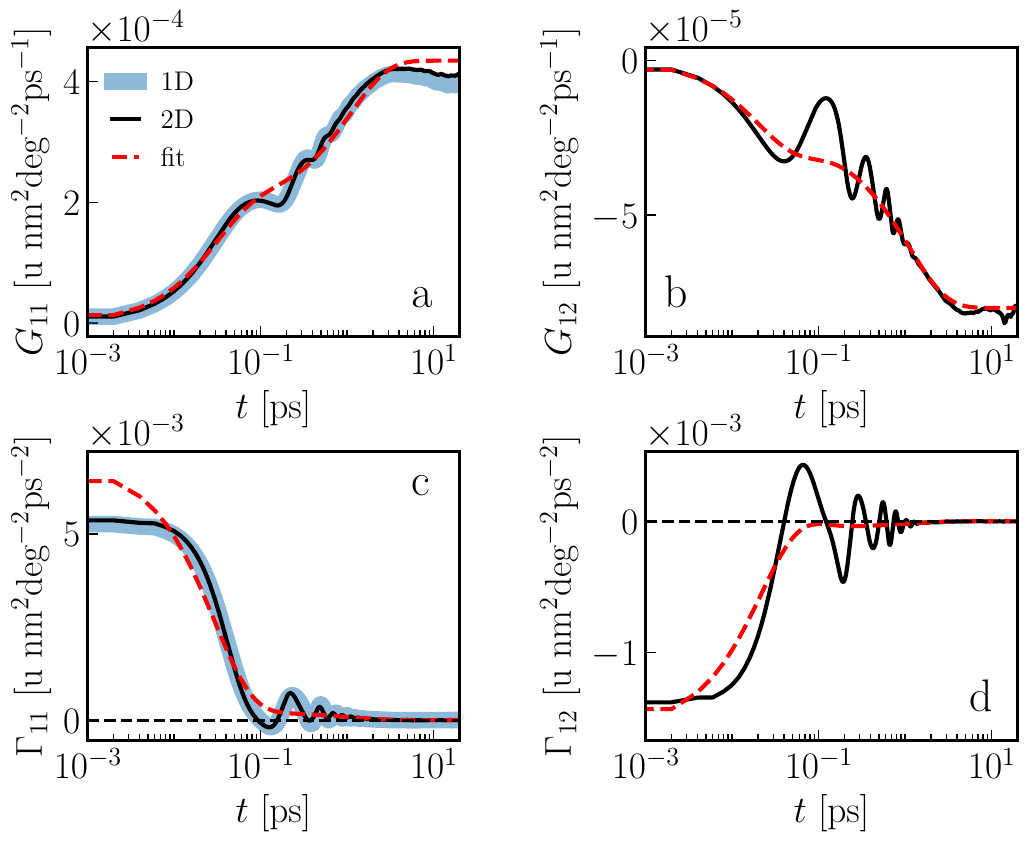}
		\caption{Extracted memory kernel matrix entries $\Gamma_{kl}$ (c, d) and running integral over the memory kernel matrix entries $G_{kl}$ (a, b) from 2D trajectories $\theta_1(t)$ (1) and $\theta_2(t)$ (2) of pentane (figure~\ref{fig:pentane_dihedrals_scheme}), computed using equation~\eqref{eq:G-iter}. Note, that we average the diagonal and off-diagonal entries (see appendix \ref{app:corr_pentane} for details). The broken lines are fits according to equation~\eqref{eq:memory_matrix} for a sum of $m=5$ matrix exponentials to the data (appendix \ref{app:fitting_procedure}). 
			The values of the fitting parameters are given in appendix \ref{app:fitting_constants_2D}. The light blue lines are 1D extraction results (section \ref{sec:mapping_1D}), where the data for $\theta_1$ and $\theta_2$ is averaged.}
		\label{fig:matrix_exp_fit_ikernels_2D_dih_pentane}
	\end{figure*}
	\begin{figure*} 
		\centering
		\includegraphics[width=1.0
		\linewidth]{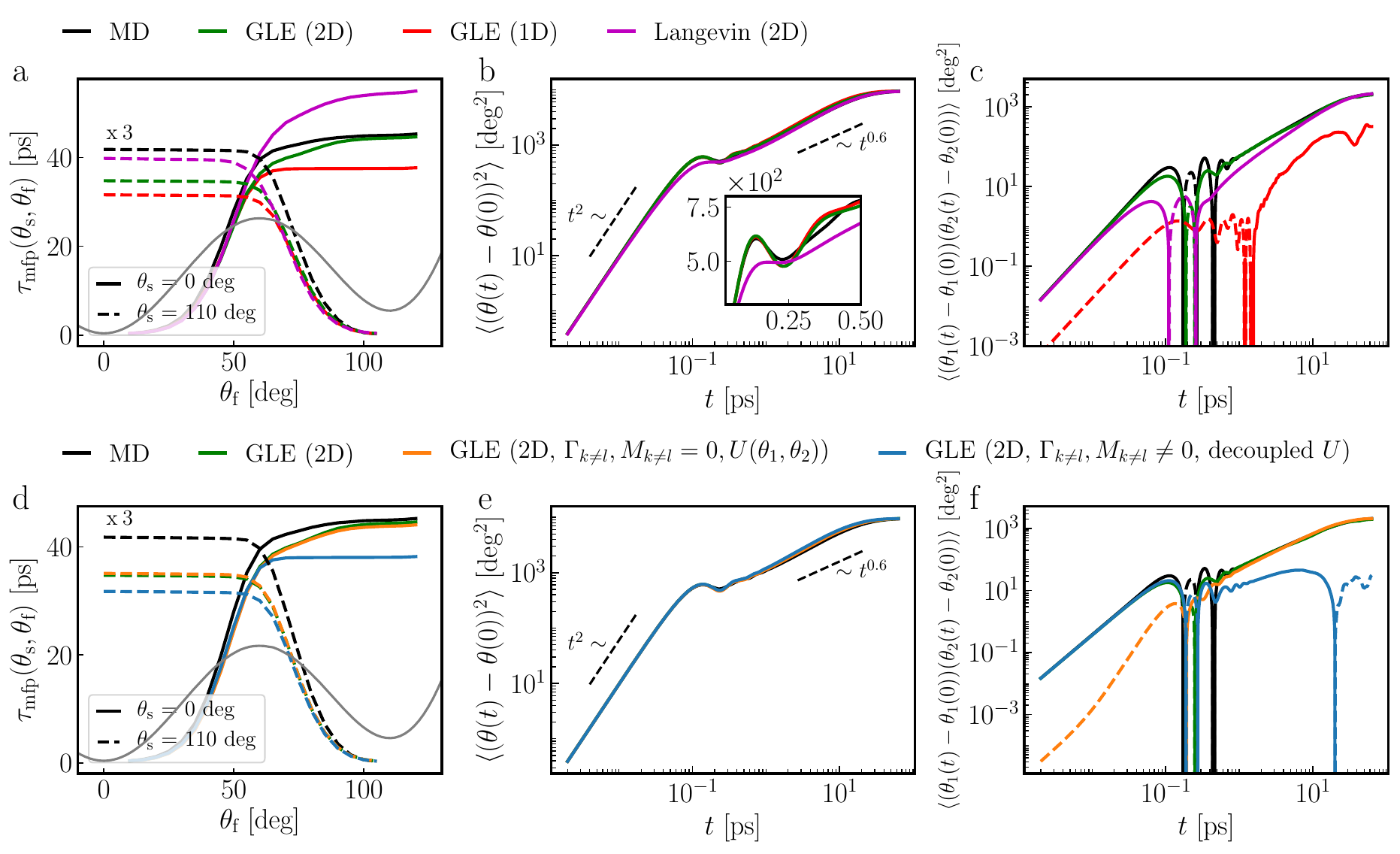}
		\caption{GLE simulation results for pentane. (a) Comparison between the mean first-passage time $\tau_\text{mfp}$ from MD (black), 1D GLE (red) and 2D GLE (green) simulations as a function of the final position $\theta_\text{f}$ for different starting positions $\theta_\text{s}$, here for the trans-to-cis transition ($\theta_\text{s}$ = 0 deg, solid lines) and cis-to-trans transition ($\theta_\text{s}$ = 110 deg, broken lines, multiplied by 3). We average over the individual results for $\theta_1$ and $\theta_2$. The 1D GLE results are computed from a simulation according to equations~\eqref{eq:markov_embedding_1D} with parameters given in appendix \ref{app:fitting_constants_1D}. The 2D GLE results correspond to a simulation according to equations~\eqref{eq:Markovian2} with parameters given in appendix \ref{app:fitting_constants_2D}. The gray curve shows the 1D potential landscape from figure~\ref{fig:pentane_dihedrals_scheme}. We compare the GLE results with Langevin simulations (purple) according to equations~\eqref{eq:Langevin2D}, where we use the fitted friction coefficients $\hat{\gamma}_i$ from the 2D memory kernel matrix given in appendix \ref{app:fitting_constants_2D}. (b, c) Mean-squared displacement $\langle(\theta(t) - \theta(0))^2\rangle$ and mean cross displacement $\langle(\theta_1(t) - \theta_1(0))(\theta_2(t) - \theta_2(0))\rangle$ from the MD and the GLE simulations in (a). The broken lines in (c) correspond to the negative values of the solid lines. The inset in (b) shows the MSD data in the sub-picoseconds range.
			(d) MFPT profiles $\tau_\text{mfp}$ from MD (black) compared with a full 2D GLE simulation (green), with a 2D GLE with diagonal matrices, i.e. $\Gamma_{k \neq l} = 0$ and $M_{k \neq l} = 0$, but 2D potential $U(\theta_1,\theta_2)$ (orange), and with a 2D GLE with decoupled potential, i.e. $U(\theta_1,\theta_2) = U_1(\theta_1) + U_2(\theta_2)$, but full $\hat{\Gamma}$ and $\hat{M}$ matrices (blue). (e, f) Mean-squared displacement and mean cross displacement from the MD and the GLE simulations in (d). The broken lines in (f) denote negative values.}
		\label{fig:mfpts_profiles_matrix_gle_sim_pentane}
	\end{figure*}
	\subsection{GLE parameters extraction}
	We apply our GLE model to describe the isomerization dynamics of a solvated pentane molecule obtained from MD simulation trajectories (see appendix \ref{app:simulation_setup} for details). In water, the pentane molecule rotates around the plane defined by the three central carbon atoms, leading to two distinct dihedral angles, $\theta_1(t)$ and $\theta_2(t)$ (see figure~\ref{fig:pentane_dihedrals_scheme}(a) for a schematic representation). Sample trajectories of these dihedral angles, along with their corresponding potential landscape $U(\theta_1,\theta_2)$ from the MD simulation, are shown in figure~\ref{fig:pentane_dihedrals_scheme}(b, c). Both dihedral angle potentials, $U_1(\theta_1)$ and $U_2(\theta_2)$, exhibit identical profiles. The angles evolve between their trans-state, i.e. $\theta_1(t),\theta_2(t) = 0$ deg, and two different cis-states, $\theta_1(t), \theta_2(t) = \pm 110$ deg. The 2D potential landscape, $U(\theta_1,\theta_2)$, reveals seven minima, suggesting that the states with distinct cis-configurations are absent. The potential, therefore, deviates from equation~\eqref{eq:additive_pmf} (appendix \ref{app:add_pot_ala2}), which can be explained by steric restrictions: the dihedral interactions in the MD simulation force field prevent both angles from simultaneously occupying different cis-states.
	\\ 
	\indent To model the pentane molecule dynamics, we combine the two reaction coordinates, $x_1(t) = \theta_1(t)$ and $x_2(t) = \theta_2(t)$ and construct a 2D GLE for $\Vec{x} = (\theta_1,\theta_2)^T$.
	In appendix \ref{app:pos_dep_mass_results}, we find that the dihedral angle of pentane exhibits a position-dependent mass. However, we reveal in appendix \ref{app:pos_dep_mass_results}, that including the position-dependent mass term in the force expression in equation~\eqref{eq:force_effective} does not result in a significant difference compared to the force derived from the potential gradient $\vec{\nabla}U(\vec{x})$. Therefore, for the following analysis, we proceed with the simplified form of the GLE in equation~\eqref{eq:multi_GLE}.
	\\
	\indent  The mass matrix, as given by equation~\eqref{eq:mass_equip}, is found to be (in units of $\text{ u nm}^{2}\text{ deg}^{-2}$)
	\begin{equation}
		\label{eq:mass_matrix_pentane}
		\Hat{M} =  \left( \begin{array}{rr} 
			2.56 \cdot 10^{-5} & -9.69 \cdot 10^{-7}  \\ 
			-9.69 \cdot 10^{-7} & 2.56 \cdot 10^{-5}  \\ 
		\end{array}\right),
	\end{equation}
	revealing that the off-diagonal mass entries are not negligible, reflecting significant cross-correlations between the velocities in the $C^{\Dot{\Vec{x}}\Dot{\Vec{x}}}$ matrix (see appendix \ref{app:corr_pentane}). In figure~\ref{fig:matrix_exp_fit_ikernels_2D_dih_pentane}, we present the results for the memory kernel matrix $\Hat{\Gamma}(t)$ and the running integral over the memory kernel matrix $\Hat{G}(t)$, obtained via equation~\eqref{eq:G-iter}. To improve statistics, we average the diagonal entries, i.e. $\Gamma_{11} = (\Gamma_{11} + \Gamma_{22})/2$, due the identical diagonal correlation functions of the dihedral angles, and the off-diagonal entries, i.e. $\Gamma_{12} = (\Gamma_{12} + \Gamma_{21})/2$, as the cross-correlation functions are identical (appendix \ref{app:corr_pentane}). The data of all entries from the extraction can be found in appendix \ref{app:entries_pentane}).
	Oscillations in the memory kernels are observed in all entries, consistent with previous findings for butane \cite{daldrop2018butane}, which we attribute to the presence of (internal) orthogonal degrees of freedom. The diagonal entry $\Gamma_{11}(t)$ is similar to 1D extraction results (light blue data, see section \ref{sec:mapping_1D}). Interestingly, $\Gamma_{12}(t)$ is negative for short and long times, with an amplitude comparable to those of the diagonal entry, implying negative friction coupling between the observables.
	\\ \indent We fit the memory kernel matrix data with the multi-exponential function defined in equation~\eqref{eq:memory_matrix}, with $m=5$ components. To reduce the number of parameters to be optimized, we assume symmetric friction coefficient matrices, i.e. $\gamma^{12}_i = \gamma^{21}_i$, and also $\gamma^{11}_i = \gamma^{22}_i$, and the same for the memory time matrices, i.e. $\tau^{11}_i = \tau^{22}_i$ and $\tau^{12}_i = \tau^{21}_i$. Applying a parallel least-squares fitting method to $\hat{\Gamma}(t)$ and $\hat{G}(t)$ simultaneously (appendix \ref{app:fitting_procedure}), we obtain the best-fitting parameters provided in appendix \ref{app:fitting_constants_2D}. The best-fitting functions for all entries are displayed as red broken lines in figure~\ref{fig:matrix_exp_fit_ikernels_2D_dih_pentane}. The observed discrepancies are expected since the oscillations in the memory kernels are neglected by the fitting function. A perfect fit can be achieved by considering exponential-oscillatory functions (appendix \ref{app:exp_osc_1D_me}).
	As given in appendix \ref{app:fitting_constants_2D}, off-diagonal memory times $\tau^{12}_i$ are not significant, which reflects the similarity between the diagonal entries in the 2D memory kernel matrix and 1D extraction results in figure~\ref{fig:matrix_exp_fit_ikernels_2D_dih_pentane}(a, c) (see appendix \ref{app:comp_kernel_matrix} for details). 
	\\ \indent For pentane, the off-diagonal entries of the friction coefficient matrices $\gamma_i^{12}$ are negative and are not negligible as the total friction coefficient, $\gamma_{12} = \sum_{i=1}^5 \gamma_i^{12}$, is approximately 18\% of the diagonal value $\gamma_{11}$ (see table \ref{tab:fit_kernel_2D_pentane_mass_01} in appendix \ref{app:fitting_constants_2D}).
	
	\subsection{Markovian embedding simulations}
	\begin{figure*}
		\centering
		\includegraphics[width=1
		\linewidth]{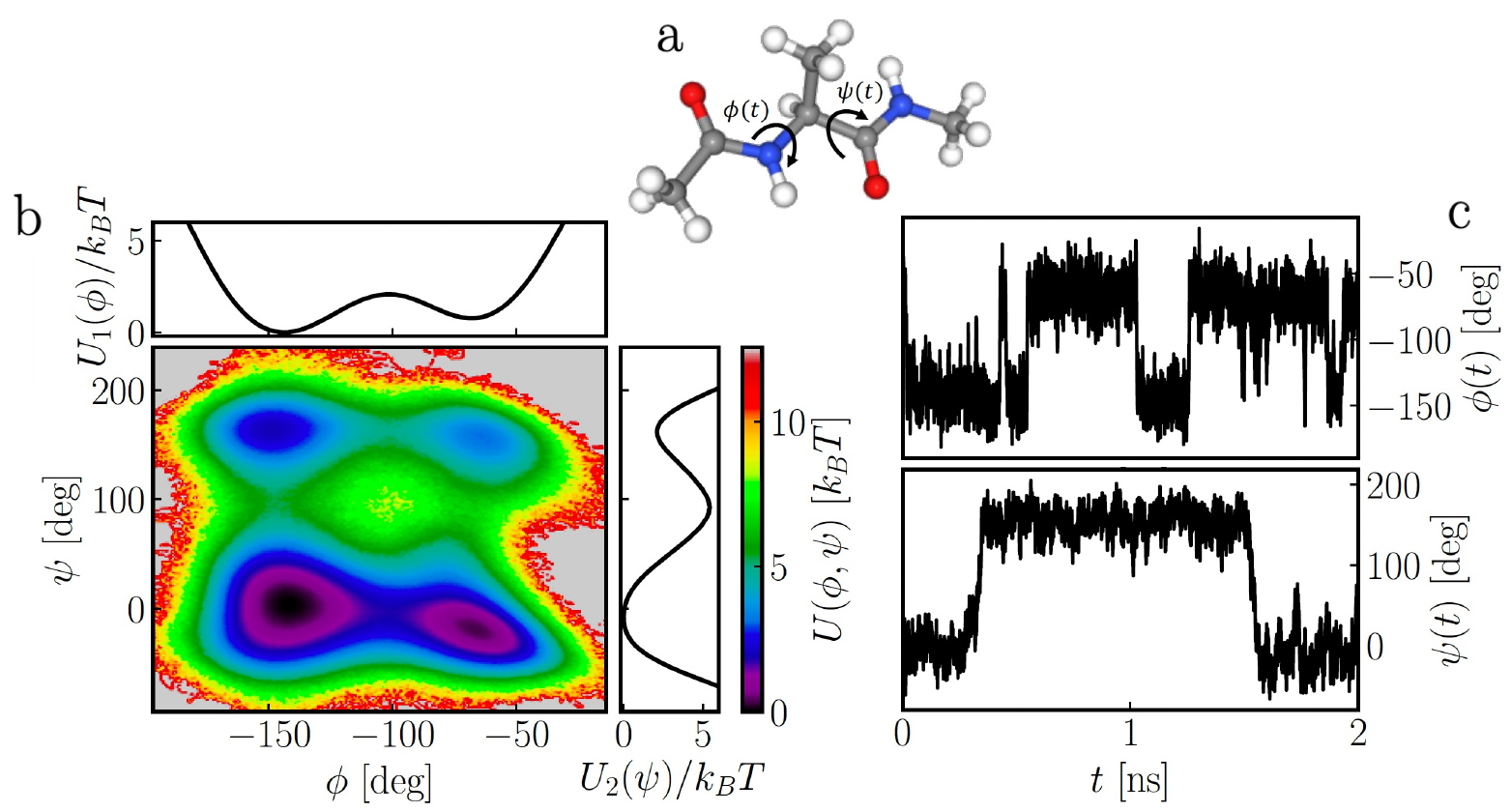}
		\caption{2D dihedral observable of alanine dipeptide. (a) Illustration of the two dihedral angles $\phi$ and $\psi$ in a alanine dipeptide molecule. (b) 2D potential $U(\phi,\psi) = -k_BT\ln \rho(\phi,\psi)$ of the dihedral angles $\phi$ and $\psi$ from a hydrated alanine dipeptide MD simulation, together with the 1D potentials $U_{1}(\phi)= -k_BT\ln \rho(\phi)$ and $U_2(\psi)= -k_BT\ln \rho(\psi)$. (c) Sample trajectories of both dihedral angles in (a).}
		\label{fig:ala2_dihedrals_scheme}
	\end{figure*}
	We perform multi-dimensional GLE simulations using the sum of matrix exponentials in equation~\eqref{eq:memory_matrix} fitted to the two-dimensional pentane dihedral angle trajectory (see section \ref{sec:mapping}). 
	For consistency, the simulation time step $\Delta t$ and the length of the GLE simulations match those of the MD simulation. In GLE simulations, we employ the effective mass matrix from equation~\eqref{eq:mass_matrix_pentane}, the 2D potential landscape shown in figure~\ref{fig:pentane_dihedrals_scheme}, and the fitted friction coefficient and memory time matrices provided in appendix \ref{app:fitting_constants_2D}.
	\\
	\indent To evaluate the predictive power of the simulations and examine the influence of various terms in the multi-dimensional GLE, we compare the mean first-passage time $\tau_\text{mfp}(\theta_\text{s},\theta_\text{f})$ (MFPT) from GLE simulations (see appendix \ref{app:simulation_setup} for details) with the MD data. The states $\theta_\text{s}$ and $\theta_\text{f}$ correspond to local minima in the 1D potential landscape shown in figure~\ref{fig:pentane_dihedrals_scheme}.
	In figure~\ref{fig:mfpts_profiles_matrix_gle_sim_pentane}(a), we display exemplary profiles of the MFPTs averaged over both dihedral angles. These profiles represent the trans-to-cis transition (starting at $\theta_\text{s}$ = 0 deg, solid lines) and the cis-to-trans transition (starting at $\theta_\text{s}$ = 110 deg, broken lines) as functions of the final position $\theta_\text{f}$. The statistical errors are calculated but smaller than the data's linewidth. 
	We compare 2D GLE simulations (green) with uncoupled 1D GLE simulations (red, see section \ref{sec:mapping_1D} for details), where the 1D potential landscapes $U_1(\theta_1)$ and $U_2(\theta_2)$ in figure~\ref{fig:pentane_dihedrals_scheme} are used. The 1D memory kernels in figure~\ref{fig:matrix_exp_fit_ikernels_2D_dih_pentane} are fitted with $m=5$ functions according to equation~\eqref{eq:fit_kernel}, and the fitting constants are provided in appendix \ref{app:fitting_constants_1D}. Appendix \ref{app:exp_osc_1D_me} includes a graphical representation of the fit.
	\\ \indent The results from the 2D GLE simulation show a slowdown compared to the 1D GLE simulation, and align more closely with the MD results (black). Notably, the improvement observed with the 2D GLE for the trans-to-cis-state transition (solid lines) is greater than for the cis-to-trans-state transition (broken lines). Note, that the 1D Markovian embedding simulation cannot be improved by a more accurate fit via exponential-oscillatory functions, as demonstrated in appendix \ref{app:exp_osc_1D_me}. Also, note, that Markovian embedding is based on the approximate relation in equation~\eqref{eq:fdt_multi}, which is generally not fulfilled for non-linear observables \cite{ayaz2022generalized, vroylandt2022gle,kiefer2025ngf}. \\
	\indent In figure~\ref{fig:mfpts_profiles_matrix_gle_sim_pentane}(b, c), we present the computed mean-squared displacement (MSD) for the MD and GLE simulations shown in (a) and mean cross displacement, which reflects the coupling between the motion of the two reaction coordinates \cite{kraft2013brownian}. The MSDs and cross displacements exhibit oscillations in the time range between 0.1 and 1 ps, indicative of memory effects \cite{mitterwallner2020negative, klimek2022optimal}. We observe subdiffusive behavior with an exponent of 0.6 for times between 1 and 10 ps, which is associated with the confinement of the reaction coordinates within a potential. 
	The MSDs from all GLE and MD simulations show excellent agreement. The mean cross displacements between the reaction coordinates from the 1D GLE in figure~\ref{fig:mfpts_profiles_matrix_gle_sim_pentane}(c) differ noticeably from the MD data, while 2D GLE results show a very good match with the MD simulations. The slight deviations between the MD and 2D GLE in the region of oscillations (0.1 - 1 ps) in (c) are expected, as the oscillations in the memory kernel matrix are neglected by the used fitting function in equation~\eqref{eq:memory_matrix}. \\
	\indent To highlight the impact of memory effects, we compare our results with those obtained from a two-dimensional Langevin equation, which corresponds to equations~\eqref{eq:Markovian2} in the limit of $\hat{\tau} \rightarrow 0$
	\begin{align}
		\label{eq:Langevin2D}
		\Hat{M} \Ddot{\Vec{x}}(t) &= -\Hat{\gamma}\Vec{\Dot{x}}(t) - \Vec{\nabla}U(\Vec{x}) + \Vec{\zeta}(t),
	\end{align}
	where $\Hat{\gamma} = \sum_{i=1}^5 \Hat{\gamma}_{i}$, and $\vec{\zeta}(t)$ represents delta-correlated Gaussian noise with $\langle \vec{\zeta}(t) \rangle = 0$ and variance
	$\langle \zeta_{k}(t) \zeta_{l}(t') \rangle =  2k_BT\Hat{\gamma}_{kl} \delta(|t-t'|)$.
	We observe a general slowdown of the Langevin simulation compared to the GLE results due to neglecting memory friction. This outcome is expected as small alkanes typically fall into the so-called memory speed-up regime \cite{kappler2018memory,kappler2019non,lavacchi2020barrier, dalton2024role}. The good agreement between the Langevin simulation and the MD data for the cis-to-trans transition in figure~\ref{fig:mfpts_profiles_matrix_gle_sim_pentane}(a) (broken lines) presumably is due to error cancellation \cite{brunig2022pair}.
	The results for the mean displacements in the range from 0.1 to 1 ps for the Langevin simulation in figure~\ref{fig:mfpts_profiles_matrix_gle_sim_pentane}(b, c) reflect that a memory-less model is not appropriate for pentane dihedral dynamics.\\
			\begin{figure*} 
		\centering
		\includegraphics[width=1
		\linewidth]{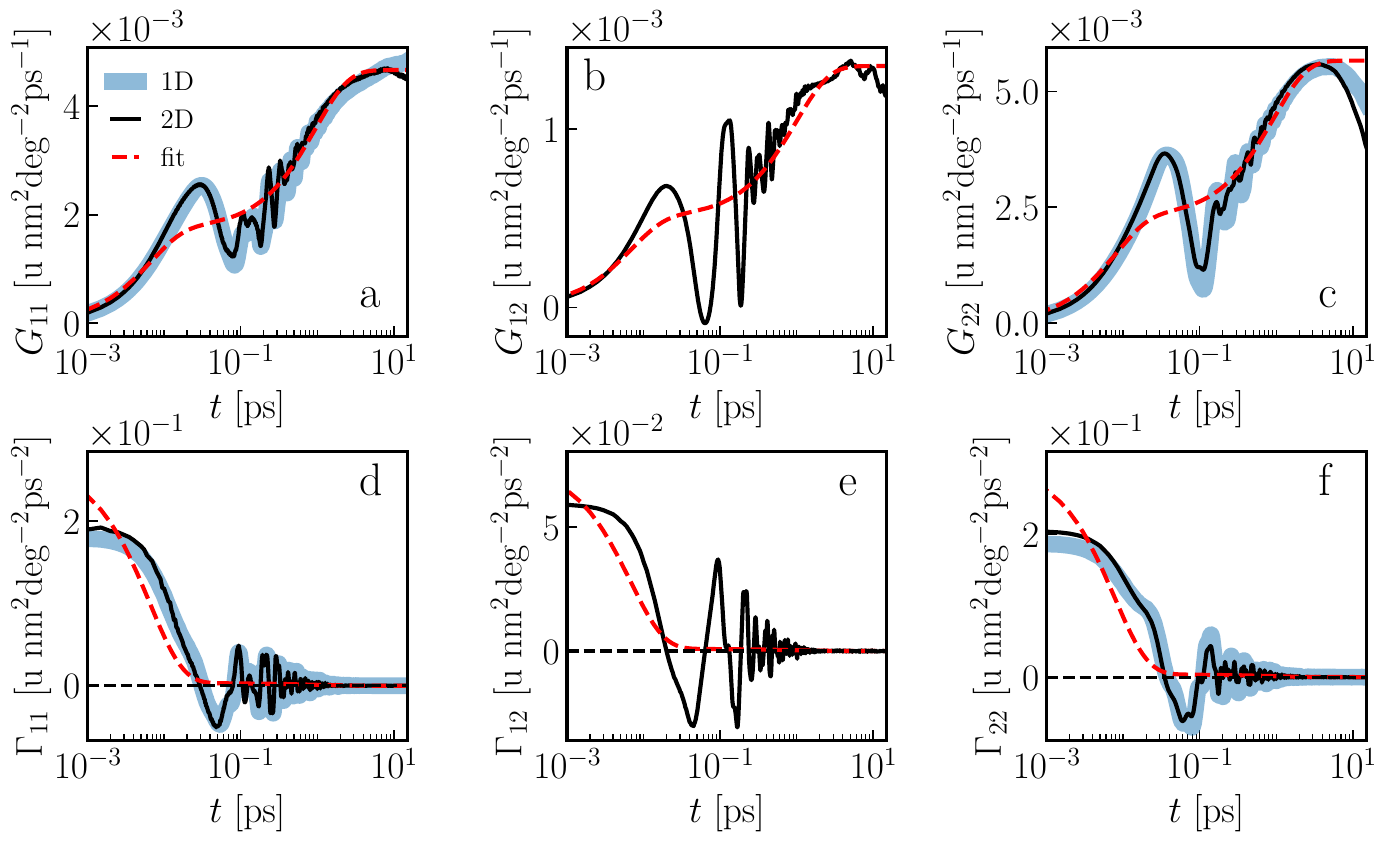}
		\caption{Extracted memory kernel matrix entries $\Gamma_{kl}$ (d - f) and running integrals over the memory kernel entries $G_{kl}$ (a - c) from two-dimensional trajectories $\phi(t)$ (1) and $\psi(t)$ (2) of alanine dipeptide, computed via equation~\eqref{eq:G-iter}. Note, that we average the off-diagonal entries. The broken lines are fits according to equation~\eqref{eq:memory_matrix} for a sum of $m=5$ matrix exponentials to the data (appendix \ref{app:fitting_procedure}). 
			The values of the fitting parameters are given in appendix \ref{app:fitting_constants_2D}. The light blue lines are 1D extraction results (section \ref{sec:mapping_1D}).}
		\label{fig:matrix_exp_fit_kernels_2D_dih_ala2}
	\end{figure*} 
	\indent We now discuss the origin of the slowdown when going from 1D to 2D GLE simulations. Potential causes include the presence of off-diagonal friction coefficients $\gamma_{12}$, off-diagonal mass entries $M_{12}$, and the multi-dimensional potential. The computed off-diagonal friction between the dihedral angles is negative and approximately a fraction of -0.18 of $\gamma_{11}$ (see figure~\ref{fig:matrix_exp_fit_ikernels_2D_dih_pentane}(b, d) and appendix \ref{app:fitting_constants_2D}), reflecting the coupling between friction of the dihedral angles. A similar result is found for the mass, with a factor of around -0.05 between the off-diagonal and diagonal entries in equation~\eqref{eq:mass_matrix_pentane}. Frictional coupling is often associated with a slowdown in the dynamics \cite{acharya2022non}, and results from the interplay between the dihedral angle and intra- or intermolecular degrees of freedom.\\
	\indent To explore the impact of off-diagonal friction on isomerization kinetics, we conduct 2D GLE simulations by neglecting individual components in equations~\eqref{eq:Markovian2}. In figure~\ref{fig:mfpts_profiles_matrix_gle_sim_pentane}(d), we compare the MFPT for the full 2D GLE (green) with a 2D GLE utilizing a decoupled potential ($U(\vec{x}) = U_1(\theta_1) + U_2(\theta_2)$, blue) and a 2D GLE with diagonal matrices ($\Gamma_{k\neq l} = 0$, $m_{k \neq l} = 0$, orange). We find that the main factor in the observed slowdown between 1D and 2D GLE simulations for pentane dynamics is the confinement of both dihedral angles within the 2D potential (green and orange). This is linked to steric restrictions, which reduce the number of accessible states from 9 to 7 minima (figure~\ref{fig:pentane_dihedrals_scheme}), limiting the configurational space and presumably slowing barrier crossing rates \cite{bryngelson1989intermediates, acharya2022non}. Our finding is further supported by the analysis of transition-path time distributions in appendix \ref{app:tpt_profiles}, where we show that only a GLE with the 2D potential accurately reproduces the transition statistics between the trans- and cis-states. \\
	\indent In figure~\ref{fig:mfpts_profiles_matrix_gle_sim_pentane}(e, f), we present the mean-squared and mean cross displacements for the simulations in (d). As observed in figure~\ref{fig:mfpts_profiles_matrix_gle_sim_pentane}(f), the full 2D GLE simulation (green) provides the best agreement with the MD data for the cross displacement. Omitting off-diagonal terms in $\hat{\Gamma}$ and $\hat{M}$ while retaining a multi-dimensional potential (orange) yields good long-time agreement beyond of $t\approx$ 1 ps, but poor short-time accuracy. In contrast, the GLE simulation with off-diagonal terms and a decoupled potential (blue) performs well at short times but shows discrepancies compared with the MD data at long times. Thus, combining a multi-dimensional potential landscape and off-diagonal coupling is crucial to accurately capture the dynamics of coupled reaction coordinates.
	\section{Conformational motion of alanine dipeptide}
	As a second application, we investigate the dihedral angle dynamics of alanine dipeptide in water, which is a classical model system for coarse-grained dynamics \cite{smith1993stochastic,hummer2003coarse,prada2009exploring,stamati2010application,leimkuhler2013robust,de2014molecular,wu2016self,kmiecik2016coarse,mardt2018vampnets}. We choose the dihedral angles $\phi(t)$ and $\psi(t)$ (see figure~\ref{fig:ala2_dihedrals_scheme}). To explore the role of off-diagonal friction and multi-dimensional landscapes, we repeat the analysis from Ref.~\cite{lee2019multi}, and analyze how the 2D GLE performs compared to decoupled 1D GLE formulations. In their study, the authors introduced a multi-dimensional GLE with a position-dependent mass matrix, and observed minimal impact of the position-dependence of the mass on the memory kernel matrix and Markovian embedding results. Hence, we adopt the multi-dimensional GLE in equation~\eqref{eq:multi_GLE} with a constant mass matrix for our analysis. Further details on the position-dependent mass are provided in appendix \ref{app:pos_dep_mass_results}.
	
	\subsection{GLE parameters extraction}
	The 2D potential landscape in figure~\ref{fig:ala2_dihedrals_scheme} includes two metastable minima for each dihedral angle, resulting in four possible states. This suggests that, unlike pentane, the decoupled composition of the 1D potential landscapes as $U(\vec{x}) = U_1(\phi) + U_2(\psi)$ is an acceptable approximation (see appendix \ref{app:add_pot_ala2} for details). 
	The mass matrix is, in units of $\text{ u nm}^{2}\text{ deg}^{-2}$, given by
	\begin{equation}
		\label{eq:mass_matrix_ala2}
		\Hat{M} =  \left( \begin{array}{rr} 
			1.28 \cdot 10^{-5}& 4.31 \cdot 10^{-6}  \\ 
			4.31 \cdot 10^{-6}  & 1.00 \cdot 10^{-5}  \\ 
		\end{array}\right).
	\end{equation}
	The diagonal entries of the memory kernel matrix $\Gamma_{11}$ and $\Gamma_{22}$  in figure~\ref{fig:matrix_exp_fit_kernels_2D_dih_ala2} are similar to 1D results (light blue data). Again, we fit the entries with a sum of $m=5$ exponential matrices, and find the off-diagonal friction coefficient entries $\gamma_{12} = \sum_{i=1}^5 \gamma_i^{12}$ to be about one-third of the diagonal values (appendix \ref{app:fitting_constants_2D}). Notably, the off-diagonal integrated memory kernel in figure~\ref{fig:matrix_exp_fit_kernels_2D_dih_ala2}(b) is positive for long times, in contrast to the results for pentane in figure~\ref{fig:matrix_exp_fit_ikernels_2D_dih_pentane}.
	
	\subsection{Markovian embedding simulations}
	\begin{figure*} 
		\centering
		\includegraphics[width=1
		\linewidth]{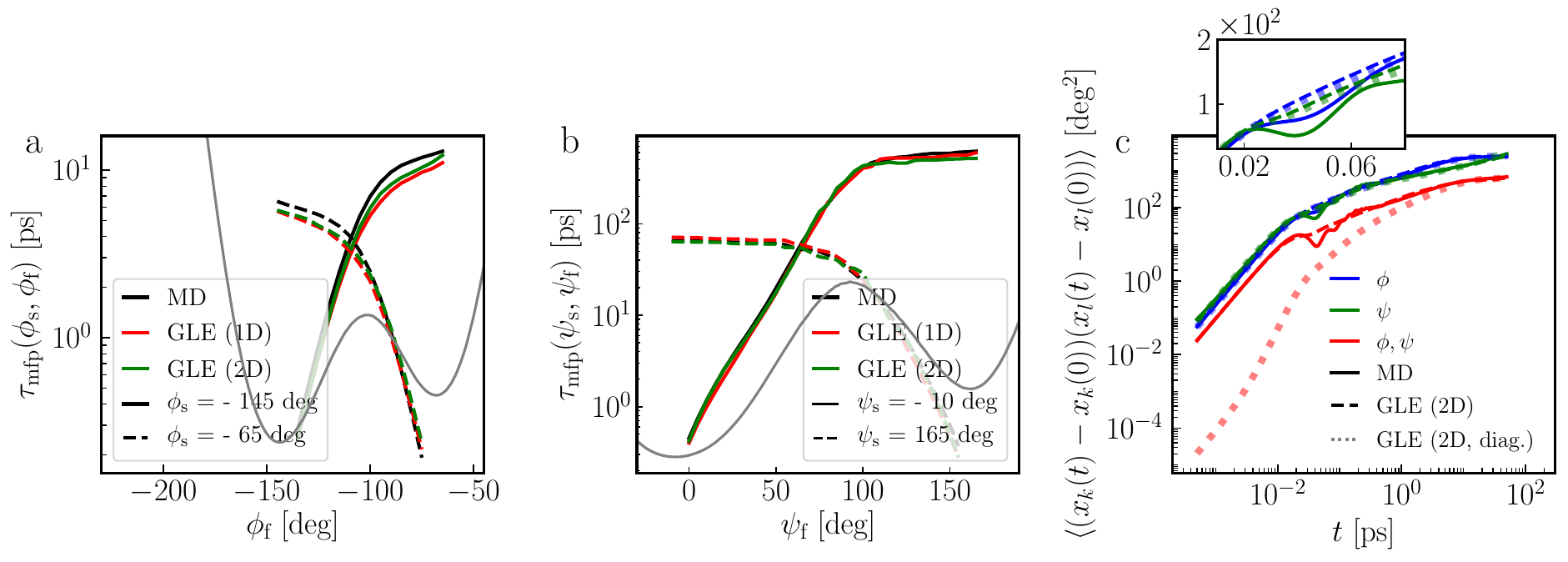}
		\caption{GLE simulation results for alanine dipeptide. (a, b) Comparison between the mean first-passage time $\tau_\text{mfp}$ from the alanine dipeptide MD (black), 1D GLE (red) and 2D GLE (green) simulations as a function of the final positions $\phi_\text{f}$ and $\psi_\text{f}$ for different starting positions $\phi_\text{s}$ and $\psi_\text{s}$, here at $\phi_\text{s}$ = -145 deg (solid lines) and $\phi_\text{s}$ = -65 deg (broken lines) for $\phi$, and at $\psi_\text{s}$ = -10 deg (solid lines) and $\psi_\text{s}$ = 165 deg (broken lines) for $\psi$. The 1D GLE results are computed from a simulation according to equations~\eqref{eq:markov_embedding_1D} with parameters given in appendix \ref{app:fitting_constants_1D}. The 2D GLE results are from a simulation according to equations~\eqref{eq:Markovian2} with parameters given in appendix \ref{app:fitting_constants_2D}. The gray curves denote the 1D potential landscapes from figure~\ref{fig:ala2_dihedrals_scheme}. (c) Mean-squared displacement (blue, green) and mean cross displacement (red) for both reaction coordinates from the MD and the 2D GLE simulations in (a, b). The dotted lines represent results from a 2D GLE simulation with diagonal matrices, i.e. $\Gamma_{k \neq l} = 0$ and $M_{k \neq l} = 0$, but using the 2D potential $U(\phi,\psi)$. Note, that we show the absolute values of the mean cross displacements. The inset shows the data in the sub-picoseconds range.}
		\label{fig:mfpts_2D_gle_sim_alanine2}
	\end{figure*} 
	Figure~\ref{fig:mfpts_2D_gle_sim_alanine2}(a, b) shows the MFPT profiles for the dihedral angles $\phi$ and $\psi$, comparing 2D GLE simulations (section \ref{sec:mapping}) with uncoupled 1D GLE simulations (section \ref{sec:mapping_1D}) employing the 1D potentials in figure~\ref{fig:ala2_dihedrals_scheme}. 
	For $\phi$, the profiles are computed starting at $\phi_\text{s}$ = -145 deg (solid lines) and $\phi_\text{s}$ = -65 deg (broken lines), while for $\psi$ they are computed starting at $\psi_\text{s}$ = -10 deg (solid lines) and $\psi_\text{s}$ = 165 deg (broken lines). We find excellent agreement between the MD and 2D GLE results for both dihedral angles. However, the agreement is already achieved with separate decoupled 1D GLEs. Thus, using a 2D GLE with off-diagonal friction coupling and the full 2D potential shows no improvement over the 1D result. \\
	\indent While Ref.~\cite{lee2019multi} modeled the dihedral angles of alanine dipeptide with a 2D GLE and found good agreement with MD first-passage time probabilities and MFPTs, they did not compare 1D and 2D GLE results to each other. Our findings suggest that, for alanine dipeptide, a 2D GLE is not advantageous over 1D GLE modeling. Moreover, we conclude that assuming sums of exponential decays for the memory kernel is sufficient to model MFPTs, and oscillations have little impact on the results, meaning that our simplified version of the Markovian embedding in Ref.~\cite{lee2019multi} is accurate.
	In contrast to pentane, the multi-dimensional confinement of the dihedral angles in alanine dipeptide does not affect the transition dynamics, shown by the fact that a 2D GLE with a 2D potential landscape (figure~\ref{fig:ala2_dihedrals_scheme}) does not improve MFPT predictions compared to the 1D GLE.\\
	\indent The diagonal MSDs in figure~\ref{fig:mfpts_2D_gle_sim_alanine2}(c, green and blue) exhibit good agreement between GLE and MD simulations. However, deviations are observed in the 0.01 to 0.1 ps range, as the GLE simulations do not capture the oscillations due to absent oscillation terms in the memory kernel fit in figure~\ref{fig:matrix_exp_fit_kernels_2D_dih_ala2}. The mean cross displacements from the GLE with purely diagonal friction (red dotted line) differ from the MD data (red solid line). In contrast, the 2D GLE including the off-diagonal memory kernels (red broken line) accurately reproduces the MD cross-correlations between the reaction coordinates. This shows that, similarly to pentane, off-diagonal friction coupling is important for correctly modeling observable correlation dynamics.
	
	
	\section{Discussion and conclusions}
	We use a multi-dimensional GLE framework to model the isomerization dynamics of solvated molecules, in conjunction with a Markovian embedding scheme that accounts for correlations between forces on reaction coordinates via off-diagonal friction coupling and confinement within a multi-dimensional potential landscape. Our GLE model effectively reproduces the dynamics observed in atomistic MD simulations of isomerization. Our study goes beyond the work of Ref.~\cite{lee2019multi} by comparing simulations between 2D GLEs and uncoupled 1D GLEs, and thereby investigating the importance of off-diagonal friction and a 2D potential.\\
	\indent The two-dimensional analysis of pentane reveals that off-diagonal memory friction between the dihedral angles is significant. Incorporating the full matrix memory friction coupling and the 2D potential landscape in the GLE simulation improves the agreement with MD results compared to 1D GLE simulations. However, the key factor in reproducing mean first-passage times is the multi-dimensional coupled potential. \\
	\indent The 2D GLE with both dihedral angles provides an accurate yet not perfect model to predict mean first-passage times, because our GLE simulation approach with Gaussian random force is approximate. As shown in appendix \ref{app:avg_pentane}, when averaging both dihedral angles, the resulting 1D GLE model with Gaussian random force also fails to capture the pentane dynamics. More accurate 1D GLE models utilizing non-Gaussian random forces or position-dependent friction could address persistent deviations between MD and GLE simulations \cite{ayaz_embedding_nl,jung2023dynamic,wolf2025cross,kiefer2025ngf}.\\
	\indent The negative off-diagonal friction in pentane, as seen in figure~\ref{fig:matrix_exp_fit_ikernels_2D_dih_pentane}(b, d), leads to positive velocity correlations (appendix \ref{app:corr_pentane}) and off-diagonal entries in the diffusion tensor $D_{kl}$ \cite{deutch1971molecular,ermak1978brownian}. In a liquid, the force on a particle is generally influenced by hydrodynamic interactions, leading to cross-correlations in the molecular displacements and torques \cite{durlofsky1987dynamic,happel2012low}. 
	This coupling has been observed in MD and Brownian dynamics simulations of n-alkanes, where transitions between neighboring dihedral angles are correlated due to angular momentum conservation and multi-dimensional confinement \cite{van1981stochastic}.
	\\
	\indent Remarkably, we reveal in appendix \ref{sec:pentane_3constr} that off-diagonal friction coupling is absent when the three inner carbon atoms in pentane are frozen, leaving only the two dihedral angles as internal degrees of freedom. This excludes hydrodynamic interactions, as hydrodynamics should not depend on whether the inner free carbon atoms are frozen or not, and shows that the off-diagonal frictional coupling of the dihedral angles involves interactions between the inner carbons' translations or rotations \cite{rosenberg1980isomerization}. Our finding highlights the significant role of intramolecular friction in pentane's conformational dynamics, offering insights without the need for viscosity-dependent experiments as in previous studies \cite{ansari1992role,daldrop2018butane, dalton2024role}.
	\\
	\indent In alanine dipeptide, contrary to pentane, positive off-diagonal friction suggests hydrodynamic interactions between the dihedral angles, presumably due to solvent-mediated effects. These interactions facilitate, rather than hinder, dihedral angle evolution. 
	While the 2D GLE captures the conformational dynamics of alanine dipeptide, it provides no clear advantage over independent 1D GLE modeling of the dihedral angles. However, including friction coupling between coordinates is essential to reproduce position cross-correlations in the form of mean cross displacements.
	\\
	\indent Our multi-dimensional GLE framework is useful for studying hydrodynamic interactions and intramolecular energy fluxes of coupled non-Markovian reaction coordinates. We demonstrate that multi-dimensional GLEs effectively capture the dynamics of complex systems, especially when forces acting on reaction coordinates are correlated due to unresolved degrees of freedom. Future studies of model systems \cite{kappler2018memory, kappler2019non, lavacchi2020barrier} with off-diagonal mass and friction coupling will help to identify scenarios for which off-diagonal memory friction significantly affects barrier crossing dynamics. 
	
		\begin{figure*}[hbt!]
		\centering
		\includegraphics[width=0.7
		\linewidth]{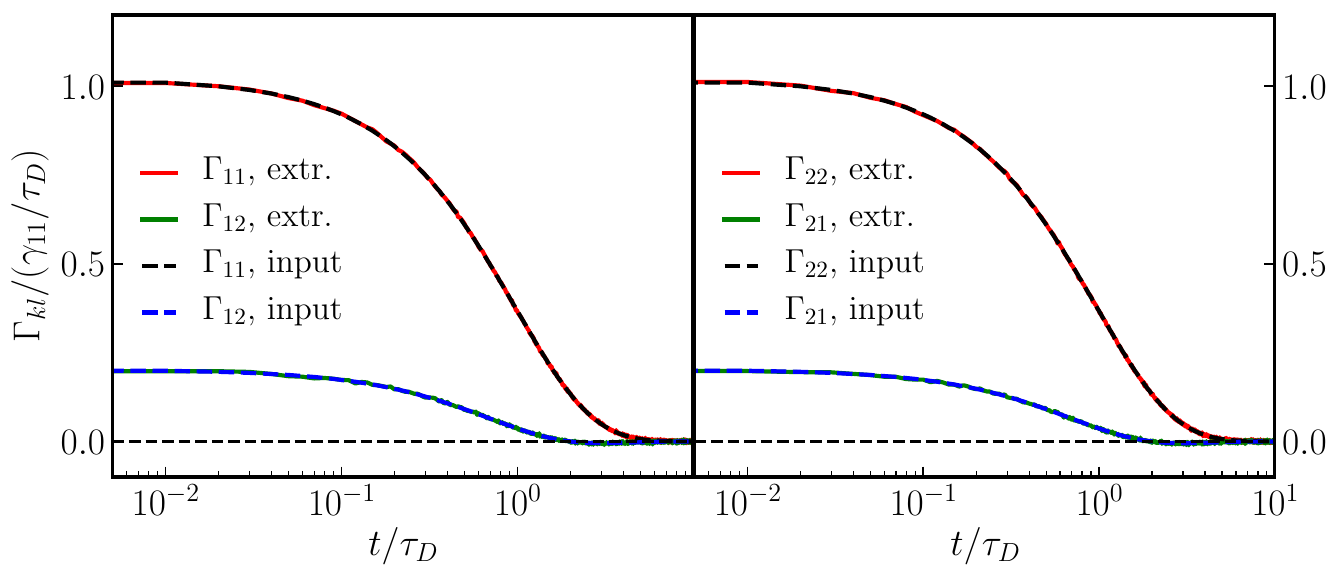}
		\caption{Extracted memory kernel matrix entries $\Gamma_{kl}$ of the two-dimensional trajectory generated from the Markovian embedding model in equations~\eqref{eq:Markovian_test}, for details see text. Here, we compare the result from the extraction method using equation~\eqref{eq:G-iter} with the analytical input, i.e. $\Hat{\Gamma}(t) = \Hat{\gamma}\Hat{\tau}^{-1}e^{-t\Hat{\tau}^{-1}}$.}
		\label{fig:2D_extraction_poc_dw}
	\end{figure*}
	
	\section*{Data availability statement}
	Input files for the MD simulations and Python code for the GLE parameter extraction and simulations, developed in this work, are available from the corresponding author upon request.
	
	\section*{Acknowledgements}
	We gratefully acknowledge financial support from Forschungsgemeinschaft (DFG) via Grant No. SFB 1449 Project Id 431232613, Project A02, and Grant No. SFB 1114 Project Id 235221301, Project C02, and from the European Research Council (ERC) under the European Union's Horizon 2020 Research and Innovation Program under Grant Agreement No. 835117. We acknowledge computing time from the HPC clusters of ZEDAT, Freie Universität Berlin.
	
	\appendix
	\section{Derivation of equation~\eqref{eq:G-iter}}
	\label{app:iter}
	Multiplying equation~\eqref{eq:multi_GLE} by $\Dot{\Vec{x}}^T(0)$ and calculating the ensemble average, we obtain a Volterra equation of the first kind
	\begin{align}
		\label{eq:multi_Volterra}
	\Hat{M} \langle \Ddot{\Vec{x}}(t) \Dot{\Vec{x}}^T(0) \rangle  =&  - \langle \vec{\nabla}U\bigl(\Vec{x}(t)\bigr) \Dot{\Vec{x}}^T(0) \rangle \\ 
		&\nonumber- \int_0^t \Hat{\Gamma}(s) \langle \Dot{\Vec{x}}(t-s) \Dot{\Vec{x}}^T(0) \rangle \:ds,
	\end{align}
	where we use $\langle\Vec{F}_\text{R}(t)\Dot{\Vec{x}}^T(0)\rangle = \Hat{0}$ \cite{ayaz2022generalized}. Equation~\eqref{eq:multi_Volterra} has a unique solution for $\hat{\Gamma}(t)$ if the autocorrelation matrix $\langle \Dot{\Vec{x}}(t-s) \Dot{\Vec{x}}^T(0) \rangle$ is invertible. We integrate equation~\eqref{eq:multi_Volterra} over time
	\begin{align}
		\label{eq:G-Volterra1}
		\Hat{M}\bigl(C^{\Dot{\Vec{x}} \Dot{\Vec{x}}}(t) - C^{\Dot{\Vec{x}} \Dot{\Vec{x}}}(0)\bigr) & \\\nonumber = - \int_0^t \:ds\: C^{\Vec{\nabla}U\Dot{\Vec{x}} }(s) &- \int_0^t \:ds\: \int_0^s \:ds'\: \Hat{\Gamma}(s') C^{\Dot{\Vec{x}} \Dot{\Vec{x}}}(s-s'),\\
		= - \int_0^t \:ds\: C^{\Vec{\nabla}U\Dot{\Vec{x}}}(s) &- \int_0^t \:ds' \int_{s'}^t \:ds\: \Hat{\Gamma}(s-s') C^{\Dot{\Vec{x}} \Dot{\Vec{x}}}(s'),\\
		= - \int_0^t \:ds\: C^{\Vec{\nabla}U\Dot{\Vec{x}}}(s) &- \int_0^t \:ds' \int_{0}^{t-s'} du\: \Hat{\Gamma}(u) C^{\Dot{\Vec{x}} \Dot{\Vec{x}}}(s'),\\
		\label{eq:G-Volterra2}
		= - \int_0^t \:ds\: C^{\Vec{\nabla}U\Dot{\Vec{x}}}(s) &- \int_0^t \:ds\: \Hat{G}(t-s) C^{\Dot{\Vec{x}} \Dot{\Vec{x}}}(s),
	\end{align}
	where $\Hat{G}(t)$ = $\int_0^t ds\:\Hat{\Gamma}(s)$ is the running integral over the memory kernel matrix $\Hat{\Gamma}(t)$, and we abbreviate $ C^{\vec{A}\vec{B}}(t) = \langle \vec{A}(t) \vec{B}^T(0) \rangle$. We further rewrite the left term of the right-hand side in equation~\eqref{eq:G-Volterra2} as
	\begin{equation}
		\int_0^t \:ds\: C^{\Vec{\nabla}U\Dot{\Vec{x}}}(s) = C^{\Vec{\nabla}U\Vec{x}}(0) - C^{\Vec{\nabla}U\Vec{x}}(t).
	\end{equation}
	Inserting this expression into equation~\eqref{eq:G-Volterra2}, we obtain the Volterra equation for the running integral
	\begin{align}
		\label{eq:G-Volterra3}
		\Hat{M} C^{\Dot{\Vec{x}} \Dot{\Vec{x}}}(t)= & \:\Hat{M} C^{\Dot{\Vec{x}} \Dot{\Vec{x}}}(0) - C^{\Vec{\nabla}U\Vec{x}}(0) + C^{\Vec{\nabla}U\Vec{x}}(t) \\\nonumber &-  \int_0^t \:ds\: \Hat{G}(t-s) C^{\Dot{\Vec{x}} \Dot{\Vec{x}}}(s). 
	\end{align}
	Discretizing this equation with a time step $\Delta t$, we derive an iterative formula for all entries of the running integral over the memory kernel matrix $G_{kl}^i$ = $G_{kl}(i\Delta t)$. For a discretized correlation matrix, we use the short-hand notation $C_i^{\Vec{A}\Vec{B}}$ = $\langle \Vec{A}(i\Delta t) \Vec{B}^T(0) \rangle$. For $\Hat{G}_i$, we obtain from equation~\eqref{eq:G-Volterra3}, by applying the trapezoidal rule on the integral,
	\begin{align}
		\label{eq:G-iter_app}
		\Hat{G}_i   &= \Bigl\lbrack C_i^{\Vec{\nabla}U\Vec{x}} - C_0^{\Vec{\nabla}U\Vec{x}} - \Hat{M}( C_i^{\Dot{\Vec{x}}\Dot{\Vec{x}}} - C_0^{\Dot{\Vec{x}}\Dot{\Vec{x}}}) \\\nonumber & - \Delta t \sum_{j=1}^{i-1}\Hat{G}_j C_{i-j}^{\Dot{\Vec{x}}\Dot{\Vec{x}}}\Bigr\rbrack \cdot (\frac{1}{2}\Delta \:  t\:C_0^{\Dot{\Vec{x}}\Dot{\Vec{x}}})^{-1}, 
	\end{align}
	where we use that $\Hat{G}_0$ = $\Hat{0}$. Extracting the potential landscape $U$($\Vec{x}$) from the given time-series data $\Vec{x}$ and applying equation~\eqref{eq:G-iter} yields the entries of the running integral $\Hat{G}(t)$ and by differentiation the memory kernel matrix $\Hat{\Gamma}(t)$.
	\section{Application of the extraction technique on a model system}
	\label{app:test_extraction}
	We test the iterative formula for the memory matrix entries $\Gamma_{kl}$ (equation~\eqref{eq:G-iter}, see section \ref{sec:extraction}) considering a 2D reaction coordinate, $\Vec{x}(t)$ = [$x_1(t)$, $x_2(t)$]$^T$. The potential is assumed to decouple according to equation~\eqref{eq:additive_pmf}: $U(\Vec{x}) = U_1(x_1) + U_2(x_2)$. The memory kernel matrix is reconstructed from simulated GLE trajectories in a double-well potential
	
	\begin{equation}
		\label{eq:double_well}
		U(\Vec{x}) = U_0 \left[\left(\frac{x_1}{L}\right)^2 - 1\right]^2 + U_0 \left[\left(\frac{x_2}{L}\right)^2 - 1\right]^2. 
	\end{equation}
	The system is described by the following extended Markovian equations
	
	\begin{align}
		\label{eq:Markovian_test}
		\Hat{M} \Ddot{\Vec{x}}(t) &= - \Vec{\nabla}U(\Vec{x}) -\Hat{\gamma}\Hat{\tau}^{-1}\bigl(\Vec{x}(t)- \Vec{y}(t)\bigr) ,\\ \nonumber
		\Dot{\Vec{y}}(t) &= - \Hat{\tau}^{-1}\bigl(\Vec{y}(t) - \Vec{x}(t)\bigr) + \Vec{\eta}(t),
	\end{align}
	which corresponds to equations~\eqref{eq:Markovian2} for a single-exponential memory kernel matrix ($m=1$ in equation~\eqref{eq:memory_matrix}). We solve these equations numerically, while choosing friction coefficients $\gamma_{12} = \gamma_{21}$ and $\gamma_{11}/\gamma_{12} = \gamma_{22}/\gamma_{12} = 10$, and a memory time of $\tau_{11}$ = $\tau_{D}$, with $\tau_{D} = L^2 \gamma_{11}/k_BT$. The off-diagonal memory time matrix entries are $\tau_{21} = \tau_{12} = -\tau_{11}/10$ and $\tau_{11} = \tau_{22}$. We use an inertial time of $\Hat{\tau}_m = \Hat{M} \Hat{\gamma}^{-1} = \Hat{\tau}$, which fixes $\Hat{M}$. The simulation is performed with a time step of $\Delta t = 0.01\tau_D$ for trajectories of length $10^6\:\cdot\:\tau_{D}$, omitting an initial equilibration time of $\tau_{D}$. Figure~\ref{fig:2D_extraction_poc_dw} verifies that the extracted memory kernel matrix matches equation~\eqref{eq:memory_matrix} within statistical error, demonstrating the robustness of our extraction technique.
	
	\section{Derivation of equation~\eqref{eq:multi_GLE} from equations~\eqref{eq:Markovian2}}
	\label{app:mapping}
	
	We derive how to construct the components of the memory kernel matrix $\hat{\Gamma}(t)$ and thereby prove the equivalence between the Markovian embedding in equations~\eqref{eq:Markovian2} and the GLE in equation~\eqref{eq:multi_GLE}. 
	To arrive at the GLE, we first solve the second equation in equations~\eqref{eq:Markovian2}
	\begin{align}
		\label{eq:sol_vec_y}
		\Vec{y}_i(t) = &\: \Vec{y}_i(0) e^{-t\Hat{\tau}_{i}^{-1}} +  \Hat{\tau}_{i}^{-1} \int_0^t dt' e^{-(t-t')\Hat{\tau}_{i}^{-1}} \Vec{x}(t') \\\nonumber & + \int_0^t dt' e^{-(t-t')\Hat{\tau}_{i}^{-1}} \Vec{\eta}_i(t').
	\end{align}
	Here, we assumed that $\Hat{\tau}_{i}$ are invertible.
	If we insert equation~\eqref{eq:sol_vec_y} into the first equation of equations~\eqref{eq:Markovian2}, we obtain
	\begin{align}
		\label{eq:Markovian21}
		\Hat{M} \Ddot{\Vec{x}}(t) = - \Vec{\nabla}U(\Vec{x}) + & \sum_{i=1}^m \Hat{\gamma}_{i}\Hat{\tau}_{i}^{-1} \Hat{\tau}_{i}^{-1}\int_0^t dt' e^{-(t-t')\Hat{\tau}_{i}^{-1}} \Vec{x}(t') \\\nonumber - \Hat{\gamma}_{i}\Hat{\tau}_{i}^{-1} \Vec{x}(t) & + \Vec{\eta}_\text{R}(t), \\
		\label{eq:Markovian22}
		= - \Vec{\nabla}U(\Vec{x}) -  \sum_{i=1}^m\Hat{\gamma}_{i}&\Hat{\tau}_{i}^{-1}  \int_0^t dt' e^{-(t-t')\Hat{\tau}_{i}^{-1}} \Dot{\Vec{x}}(t') + \Vec{\eta}_\text{R}(t).
	\end{align}
	In equation~\eqref{eq:Markovian22}, we have used integration by parts and we define
	\begin{align}
		\Vec{\eta}_\text{R}(t) =  &\sum_{i=1}^m \Vec{\eta}_i^R(t) =\sum_{i=1}^m \Hat{\gamma}_{i}\Hat{\tau}_{i}^{-1} \Vec{y}_i(0) e^{-t\Hat{\tau}_{i}^{-1}}   \\\nonumber & + \Hat{\gamma}_{i}\Hat{\tau}_{i}^{-1} \int_0^t dt' e^{-(t-t')\Hat{\tau}_{i}^{-1}} \Vec{\eta}_i(t').
	\end{align}
	To obtain the equivalence between equations~\eqref{eq:Markovian2} and equation~\eqref{eq:multi_GLE} and the validity of equation~\eqref{eq:fdt_multi}, we need to show that $\Vec{\eta}_\text{R}(t) = \sum_{i=1}^m\Vec{\eta}_i^R(t)$ follows a stationary Gaussian process with the same first moments as $\Vec{F}_\text{R}(t)$, i.e. $\Vec{\eta}_\text{R}(t)  \equiv$  $\Vec{F}_\text{R}(t)$ \cite{kappler2019non}. 
	For this, we assume that the initial conditions $\Vec{y}(0)$ are Gaussian variables with zero mean and variance
	\begin{align}
		\langle \Vec{y}_i(0) \Vec{y}_j^T(0) \rangle &= \delta_{ij} k_BT \Hat{\tau}_{i}\Hat{\gamma}_{i}^{-1},\\
		\langle y_{i}^{kl}(0) y_{i}^{kn}(0) \rangle &= \delta_{kl}\delta_{kn} k_BT \Hat{\tau}_{i}^{kl}(\Hat{\gamma}_{i}^{kl})^{-1},
	\end{align}
	and furthermore that $\langle \Vec{y}_i(0) \Vec{\eta}_j^T(0) \rangle $ = 0. 
		\begin{figure*} [hbt!]
		\centering
		\includegraphics[width=0.75
		\linewidth]{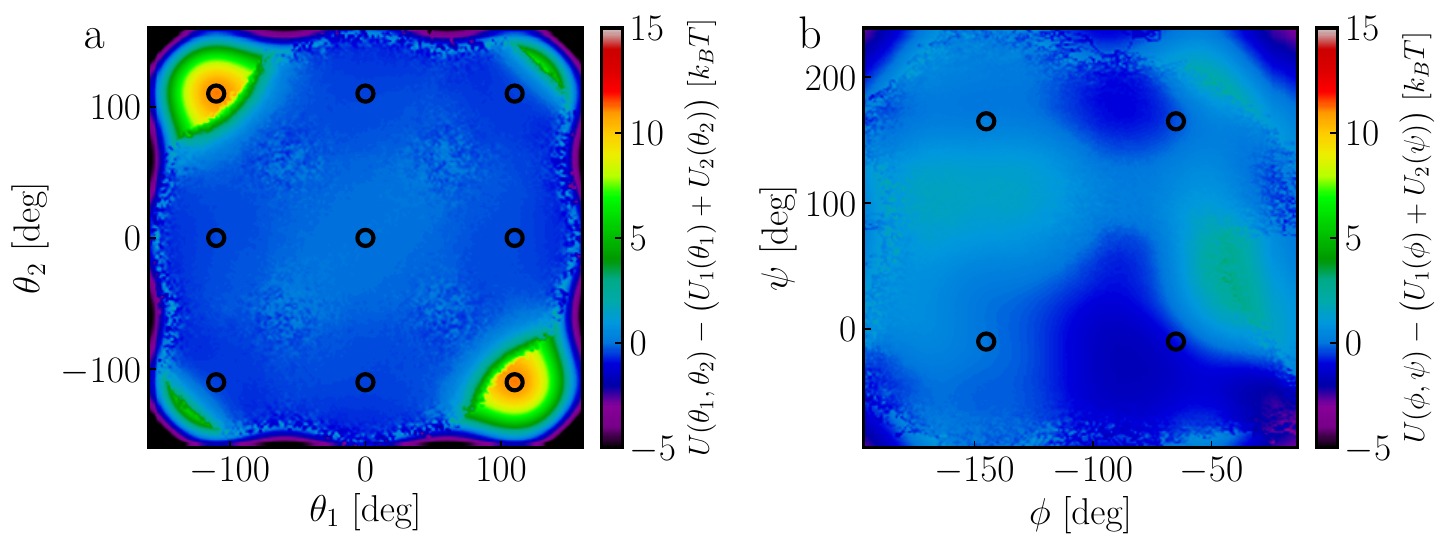}
		\caption{(a) Difference between the 2D potential $U(\theta_1,\theta_2)$ of pentane, shown in figure~\ref{fig:pentane_dihedrals_scheme}, and the decoupled potential $U_1(\theta_1) + U_2(\theta_2)$. $U_1(\theta_1)$ and $U_2(\theta_2)$ are shown in figure~\ref{fig:pentane_dihedrals_scheme}(b). The black circles denote the minima of $U_{1,2}$.
			(b) Difference between the 2D potential $U(\phi,\psi)$ of alanine dipeptide, shown in figure~\ref{fig:ala2_dihedrals_scheme}, and the decoupled potential $U_1(\phi) + U_2(\psi)$. $U_1(\phi)$ and  $U_2(\psi)$ are shown in figure~\ref{fig:ala2_dihedrals_scheme}(b).}
		\label{fig:add_pot_ala2}
	\end{figure*}  
	The mean of $\Vec{\eta}_\text{R}(t)$ is easily seen to be zero, and the variance is given by (utilizing equation~\eqref{eq:Markovian3})
	\begin{widetext}
		\begin{align}
			\label{eq:B6}
			\langle \Vec{\eta}_\text{R}(t) \Vec{\eta}_\text{R}^T(t') \rangle &= k_BT \sum_{i=1}^m \Hat{\gamma}_{i}\Hat{\tau}_{i}^{-1} e^{-(t+t')\Hat{\tau}_{i}^{-1}} + 2k_BT \sum_{i=1}^m \Hat{\gamma}_{i}\Hat{\tau}_{i}^{-1}\Hat{\gamma}_{i}\Hat{\tau}_{i}^{-1} \int_0^t du \int_0^{t'} du'\: e^{-(t+t'-u-u')\Hat{\tau}_{i}^{-1}}\Hat{\gamma}_{i}^{-1}\delta(u-u'),
			\\ \label{eq:B7}
			&=  k_BT \sum_{i=1}^m \Hat{\gamma}_{i}\Hat{\tau}_{i}^{-1} e^{-(t+t')\Hat{\tau}_{i}^{-1}} + 2k_BT \sum_{i=1}^m \Hat{\gamma}_{i}\Hat{\tau}_{i}^{-1}\Hat{\gamma}_{i}\Hat{\tau}_{i}^{-1} \int_0^{\text{min}\{t,t'\}} du \: e^{-(t+t'-2u)\Hat{\tau}_{i}^{-1}}\Hat{\gamma}_{i}^{-1},\\
		 \label{eq:B8}
			&=  k_BT \sum_{i=1}^m \Hat{\gamma}_{i}\Hat{\tau}_{i}^{-1} e^{-(t+t')\Hat{\tau}_{i}^{-1}} + k_BT \sum_{i=1}^m \Hat{\gamma}_{i}\Hat{\tau}_{i}^{-1} \bigl(e^{-(t+t' - 2\text{min}\{t,t'\})\Hat{\tau}_{i}^{-1}} - e^{-(t+t')\Hat{\tau}_{i}^{-1}}\bigr),
		\end{align}	\end{widetext}
	which results to
\begin{align} 
		\label{eq:B9}
		\langle \Vec{\eta}_\text{R}(t) \Vec{\eta}_\text{R}^T(t') \rangle  = k_BT \sum_{i=1}^m  \Hat{\gamma}_{i}\Hat{\tau}_{i}^{-1} e^{-(|t-t'|)\Hat{\tau}_{i}^{-1}}.
		\end{align}
Here, we use that $t+t' - 2\text{min}\{t,t'\} = |t-t'|$. 
	We find that $\Vec{\eta}_\text{R}(t) \equiv$  $\Vec{F}_\text{R}(t)$ and equations~\eqref{eq:Markovian2} and equation~\eqref{eq:multi_GLE} are equivalent if the memory kernel matrix entries have the form in equation~\eqref{eq:memory_matrix}.
Note, that the derivation also holds for asymmetric friction coefficient matrices and symmetric memory times matrices, i.e. $\gamma_i^{kl} \neq \gamma_i^{lk}$ and $\tau_i^{kl} = \tau_i^{lk}$, since the identity $\gamma_i \tau_i^{-1}\tau_i\gamma_i^{-1} = \mathbf{1}$ we used to arrive from equation~\eqref{eq:B7} to equation~\eqref{eq:B8} also holds in this more general case.

\section{MD simulation details}
\label{app:simulation_setup}
\begin{figure*} 
	\centering
	\includegraphics[width=1
	\linewidth]{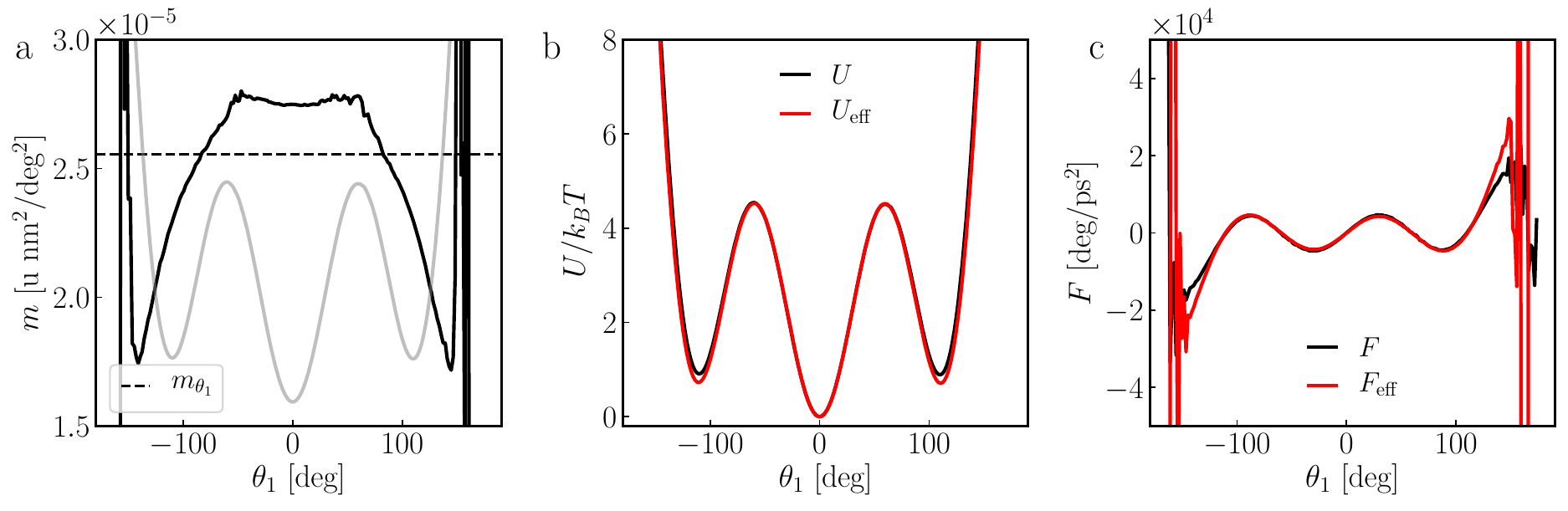}
	\caption{(a) Extracted position-dependent mass $m(\theta_1)$ = $k_BT/\langle \Dot{\theta}_1^2\rangle_{\theta_1}$ from the pentane MD simulation in figure~\ref{fig:pentane_dihedrals_scheme} for the dihedral angle $\theta_1$. The broken line denotes the constant mass, i.e. $m_{\theta_1} = k_BT / \langle \Dot{\theta}_1^2 \rangle $. The gray curve shows the PMF in (b). (b) Comparison between the PMF $U(\theta_1)$ (black), and the effective potential with mass correction, i.e. $U_\text{eff}(\theta_1) = U(\theta_1) + k_BT \ln m(\theta_1)$ (red). (c) Comparison between the mean force without, i.e. $F(\theta_1) = \nabla U(\theta_1)/m_{\theta_1}$, and with mass correction, i.e. $F_{\text{eff}}(\theta_1) = \nabla \bigl(U(\theta_1) + k_BT \ln m(\theta_1)\bigr)/m(\theta_1)$, which follows from equation~\eqref{eq:force_effective}.}  
	\label{fig:2D_dih_pentane_pos_dep_mass}
\end{figure*}  

\begin{figure*} 
	\centering
	\includegraphics[width=1
	\linewidth]{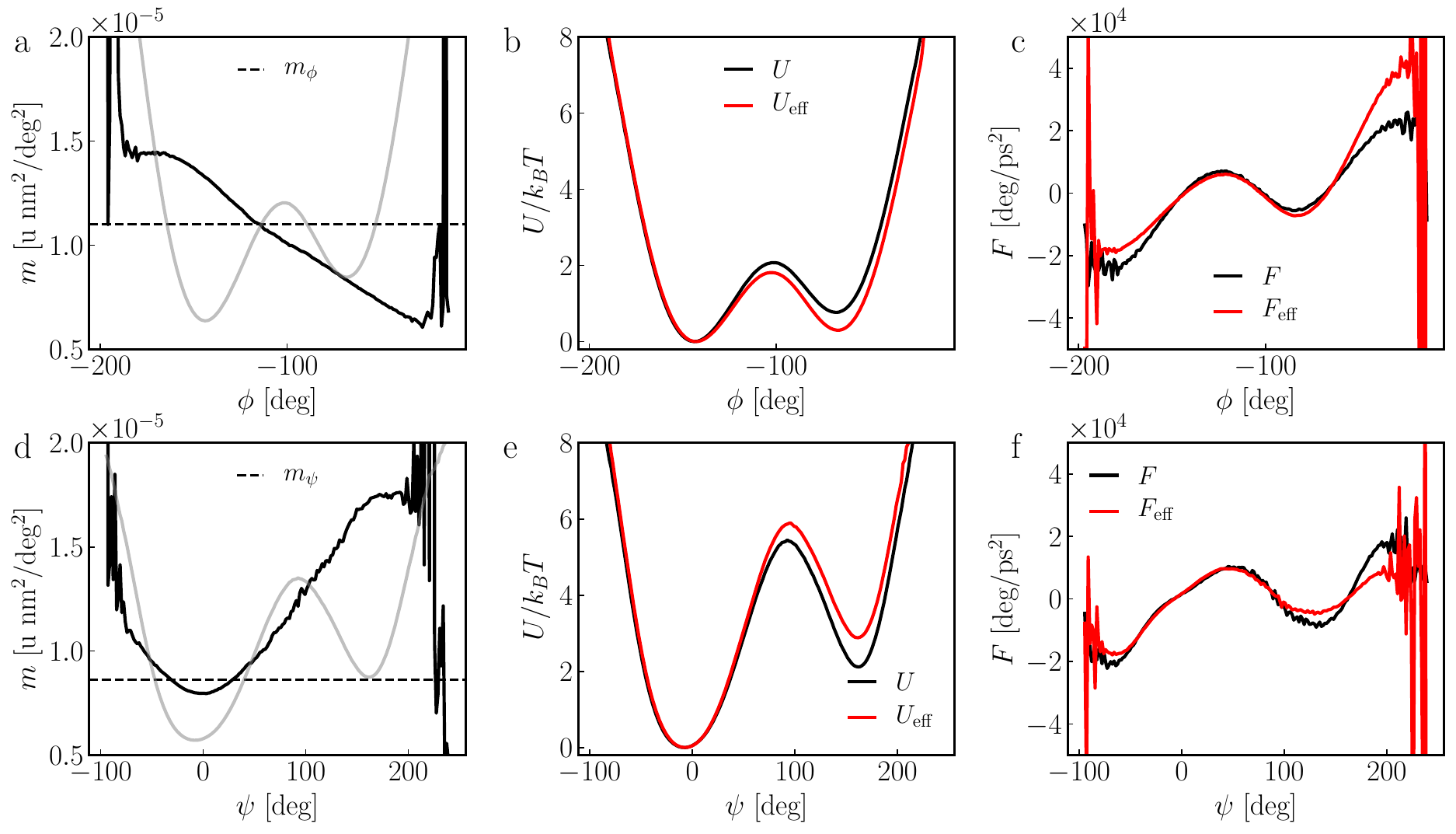}
	\caption{Same analysis as in figure~\ref{fig:2D_dih_pentane_pos_dep_mass}, here for alanine dipeptide (compare figure~\ref{fig:ala2_dihedrals_scheme}), (a - c) for $\phi$ and (d - f) for $\psi$.}
	\label{fig:2D_dih_diala_pos_dep_mass}
\end{figure*}

\begin{figure*}
	\centering
	\includegraphics[width=0.7
	\linewidth]{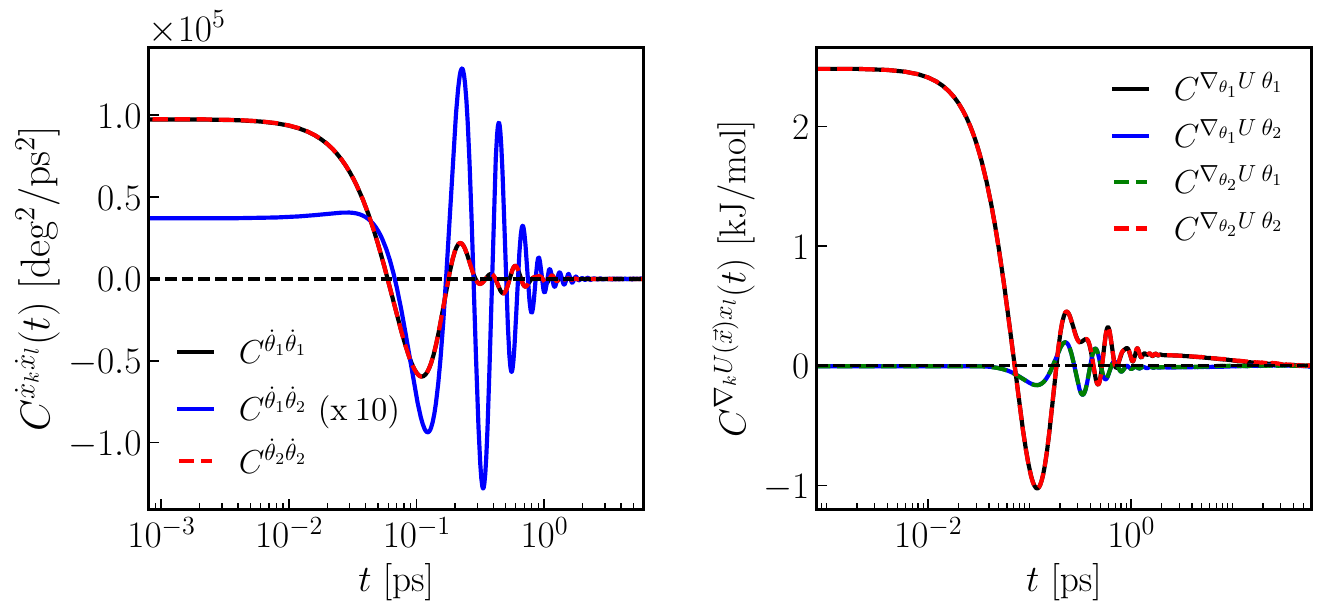}
	\caption{Entries of the 2D velocity-correlation matrix $C^{\Dot{x}_k \Dot{x}_l}(t)$ and entries of the 2D force-position-correlation matrix $C^{\nabla_k U(\Vec{x})x_l}(t)$, computed from the dihedral angle trajectories of pentane ($\theta_1$ and $\theta_2$, see figure~\ref{fig:pentane_dihedrals_scheme}).}
	\label{fig:corrs_pentane_2D}
\end{figure*}    
We perform MD simulations utilizing the GROMACS software package (version 2021.3) \cite{pronk2013gromacs}.
For the MD simulations of pentane, we place a single pentane molecule in a (4.1 nm)$^3$ water box, using the GROMOS united-atom force field \cite{Oostenbrink_2004}, which represents pentane molecule as five Lennard-Jones beads with fixed bond lengths and angles (SHAKE algorithm \cite{Ryckaert_1977}). The 1000 water molecules are modeled with the SPC/E model \cite{berendsen1987missing}, and the system is pre-equilibrated in an NPT ensemble at pressure $P$ = 1 atm and temperature $T$ = 300 K with a Berendsen barostat \cite{Berendsen_1984}. Production runs are performed in the NVT ensemble with a total simulation length of 1 $\mu$s and a time step of $\Delta t$ = 2 fs. The temperature $T$ is maintained at 300 K via the velocity rescaling thermostat \cite{bussi2007canonical}. Electrostatics are handled with the particle-mesh Ewald method \cite{Darden_1993} and a cut-off of 1 nm.\\
\indent Additionally, we simulate a solvated alanine dipeptide molecule in a cubic (1.2 nm)$^3$ water box , using the Amber99 force field for the peptide \cite{salomon2013overview} and the TIP3P water model \cite{jorgensen1983comparison} for the 1134 water molecules. Bond lengths and angles are fixed by the LINCS algorithm \cite{hess1997lincs}. The NVT simulations at $T$ = 300 K are run for 250 ns with a time step of $\Delta t$ = 0.5 fs, following the settings in Ref.~\cite{lee2019multi}.\\
\indent The mean-squared and cross displacement from MD and GLE simulations are computed using the \textit{tidynamics} package \cite{de2018tidynamics}. The mean first-passage time, $\tau_\text{mfp}$($x_\text{s}$, $x_\text{f}$), is calculated from trajectories as the average time to go from initial state $x_\text{s}$ to final state $x_\text{f}$. For this, the first-passage time (FPT) distribution is obtained from the simulations, thereby considering all FPT events including recrossings of the initial state $x_\text{s}$ \cite{dalton2024role}.

\section{2D potential landscapes of pentane and alanine dipeptide}
\label{app:add_pot_ala2}
In figure~\ref{fig:add_pot_ala2}(a), we show the difference between the total potential $U(\theta_1,\theta_2)$ of the dihedral angles of pentane and the decoupled potential $U_1(\theta_1) + U_2(\theta_2)$. Overall, the difference is small. The only exception are the states (black circles) with different cis-configurations; this verifies that the decoupled potential in equation~\eqref{eq:additive_pmf} is not a good approximation of the potential $U(\theta_1,\theta_2)$.\\
\indent Contrary, the lack of additional minima beyond those of $U_1$ and $U_2$ for the dihedral angles potential  of alanine dipeptide (figure~\ref{fig:ala2_dihedrals_scheme}(b)) supports a decoupled approximation for the 2D potential in equation~\eqref{eq:additive_pmf}. In figure~\ref{fig:add_pot_ala2}(b), we plot the difference between the total potential $U(\phi,\psi)$ of alanine dipeptide and the decoupled potential $U_1(\phi) + U_2(\psi)$. Particularly at the minima (black circles), the difference is small. However, we observe regions where the difference is non-zero and exceeds 1 $k_BT$, particularly at the $\psi = 90$ deg barrier and near the right minimum of $\phi$ around 50 deg, presumably due to the sampling from finite length MD trajectories.
These findings indicate that while the decoupled Ansatz for the 2D potential is not exact, it serves as a reasonable approximation for alanine dipeptide.

\section{Position-dependent mass of the dihedral angles of pentane and alanine dipeptide}
\label{app:pos_dep_mass_results}

The generalized mass of the pentane dihedral angle $\theta_1(t)$, given by $m(\theta_1) = k_BT/\langle \Dot{\theta}_1^2\rangle_{\theta_1}$, is weakly position-dependent, as demonstrated in figure~\ref{fig:2D_dih_pentane_pos_dep_mass}(a); the same was observed for butane in Ref.~\cite{ayaz2022generalized}. This results in a correction of the mean force term in equation~\eqref{eq:force_effective}. However, in (b) and (c), the modified potential $U_\text{eff}(\theta_1) = U(\theta_1) + k_BT \ln m(\theta_1)$ and the effective force $F_{\text{eff}}(\theta_1) = \nabla \bigl(U(\theta_1) + k_BT \ln m(\theta_1)\bigr)/m(\theta_1)$ show only small differences from the PMF $U(\theta_1) = - k_BT \ln \rho(\theta_1)$. Therefore, extracting the memory kernel from a GLE with position-dependent mass (equation~\eqref{eq:multi_GLE_hybrid}) is expected to not significantly alter the results presented in the main text.\\
\indent As found in figure~\ref{fig:2D_dih_diala_pos_dep_mass}, position-dependent mass effects on the potential and effective force are more significant for alanine dipeptide, especially at the right minimum of $U$ for both dihedral angles (b, e). However, these effects do not significantly influence memory kernel extraction or GLE simulation, as shown in Ref.~\cite{lee2019multi}.

\section{Correlation matrices for pentane}
\label{app:corr_pentane}

In figure~\ref{fig:corrs_pentane_2D}, we depict the cross-correlation and autocorrelation functions for the two-dimensional dihedral angle observable of pentane. The velocity autocorrelations $C^{\Dot{\theta}_1 \Dot{\theta}_1}(t)$ and $C^{\Dot{\theta}_2 \Dot{\theta}_2}(t)$ are, as expected, identical, decaying over equal time scales with the same amplitude. The velocity cross-correlation $C^{\Dot{\theta}_1 \Dot{\theta}_2}(t)$ decays similarly, with a positive amplitude. The cross-correlations between position and force along different reaction coordinates (blue, green) have vanishing values at $t=0$ and are identical. Hence, extracting the memory kernel matrix $\hat{\Gamma}(t)$ according to equation~\eqref{eq:G-iter} yields similar diagonal and off-diagonal entries; this is why, for pentane, we average them in the main text.

\section{Fitting procedure of the memory kernel matrix}
\label{app:fitting_procedure}
Fitting the memory kernel matrix $\hat{\Gamma}(t)$ typically requires explicit functions for all entries (see appendix \ref{app:comp_kernel_matrix}), which become complex for higher dimensions. To simplify, we fit all entries in parallel. First, we fit the diagonal entries with one-dimensional single-exponentials according to equation~\eqref{eq:fit_kernel}, then we use these as initial values for a full multi-exponential matrix fit with equation~\eqref{eq:memory_matrix}. The Levenberg-Marquardt algorithm in \textit{scipy} (version 1.4) \cite{2020SciPy-NMeth} optimizes the parameters. Our method can be easily extended to higher dimensions without writing out explicit entries from equation~\eqref{eq:memory_matrix}. We constrain the parameter space for diagonal entries to positive values and for off-diagonal entries to be smaller than the diagonal entries, ensuring invertibility. The data set is filtered beforehand on a logarithmic time scale to reduce the overall number of data points to fit. In the full fit, for the one-dimensional data as well as for the two-dimensional data, we optimize the parameters such that the fitting residuals for the memory kernel matrix $\Hat{\Gamma}(t)$ and the running integral $\Hat{G}(t)$ become minimal simultaneously. This allows us to fit the short- and long-time regimes. 
The fitting parameters are summarized in appendix \ref{app:fitting_constants_2D} and \ref{app:fitting_constants_1D}.

\section{Fitting constants of the two-dimensional memory kernel fits}
\label{app:fitting_constants_2D}
This appendix summarizes the 2D memory kernel matrix fitting parameters in tables \ref{tab:fit_kernel_2D_pentane_mass_01} and \ref{tab:fit_kernel_2D_ala2}.
\begin{table}[!htbp]
	\centering
	\caption{Fitting parameters for the memory kernel matrix fit according to equation~\eqref{eq:memory_matrix} for the pentane data shown in figure~\ref{fig:matrix_exp_fit_ikernels_2D_dih_pentane}.  We assume symmetric friction coefficient matrices, i.e. $\gamma_i^{11} = \gamma_i^{22}$ and $\gamma_i^{12} = \gamma_i^{21}$ and symmetric memory time matrices, i.e. $\tau_i^{11} = \tau_i^{22}$ and $\tau_i^{12} = \tau_i^{21}$. The units for $\gamma_i^{kl}$ are 10$^{-4}$ u nm$^{2}$ deg$^{-2}$ ps$^{-1}$ and for  $\tau_i^{kl}$ ps.}
	\begin{ruledtabular}
		\begin{tabular}{c  c  c  c  c  } 
			$i$ & $\gamma_i^{11}$ &  $\gamma_i^{12}$  & $\tau_i^{11}$ &  $\tau_i^{12}$ \\
			\hline
			1 & 0.01 &  -0.006 & 1.14 & 0.001 \\
			2 & 0.01 & -0.006 & 1.19 & 0.001 \\
			3 & 0.01 & -0.006 & 1.17 & 0.001 \\
			4 & 1.97 & -0.26 & 0.03 & 0.003 \\
			5 & 2.35 &  -0.52 & 1.11 & 0.001 \\  
			$\sum_{i=1}^5 \gamma_i^{kl}$ & 4.35 & -0.80  &  &   \\
		\end{tabular}
		
	\end{ruledtabular}
	\label{tab:fit_kernel_2D_pentane_mass_01}
\end{table}

\begin{table}[!htbp]
	\centering
	\caption{Fitting parameters for the memory kernel matrix fit according to equation~\eqref{eq:memory_matrix} for the alanine dipeptide data shown in figure~\ref{fig:matrix_exp_fit_kernels_2D_dih_ala2}. For the fitting, we assume $\gamma_i^{12} = \gamma_i^{21}$ and $\tau_i^{12} = \tau_i^{21}$. The units for $\gamma_i^{kl}$ are 10$^{-3}$ u nm$^{2}$ deg$^{-2}$ ps$^{-1}$ and for  $\tau_i^{kl}$ ps.}
	\begin{ruledtabular}
		
		\begin{tabular}{c  c  c  c  c  c  c} 
			
			$i$ & $\gamma_i^{11}$ & $\gamma_i^{22}$ &  $\gamma_i^{12}$  & $\tau_i^{11}$ & $\tau_i^{22}$&  $\tau_i^{12}$ \\
			\hline
			1 & 0.01 & 0.01 &  0.001 & 1.74 & 1.28 & 0.001 \\
			2 & 0.01 & 0.27 & 0.001 & 0.95 & 0.92 & 0.001 \\
			3 & 0.29 & 0.28 & 0.09  & 0.94 & 0.92 & 0.001 \\
			4 & 1.73 & 2.28 & 0.50  & 0.01 & 0.01 & 0.001 \\
			5 & 2.63 & 2.83 & 0.76 & 0.95 & 0.92 & 0.001 \\
			$\sum_{i=1}^5 \gamma_i^{kl}$ & 4.67 & 5.66  & 1.35  &  &  &  \\
		\end{tabular}
		
	\end{ruledtabular}
	\label{tab:fit_kernel_2D_ala2}
\end{table}

\section{Memory kernel entries for pentane}
\label{app:entries_pentane}
\begin{figure*} 
	\centering
	\includegraphics[width=0.7
	\linewidth]{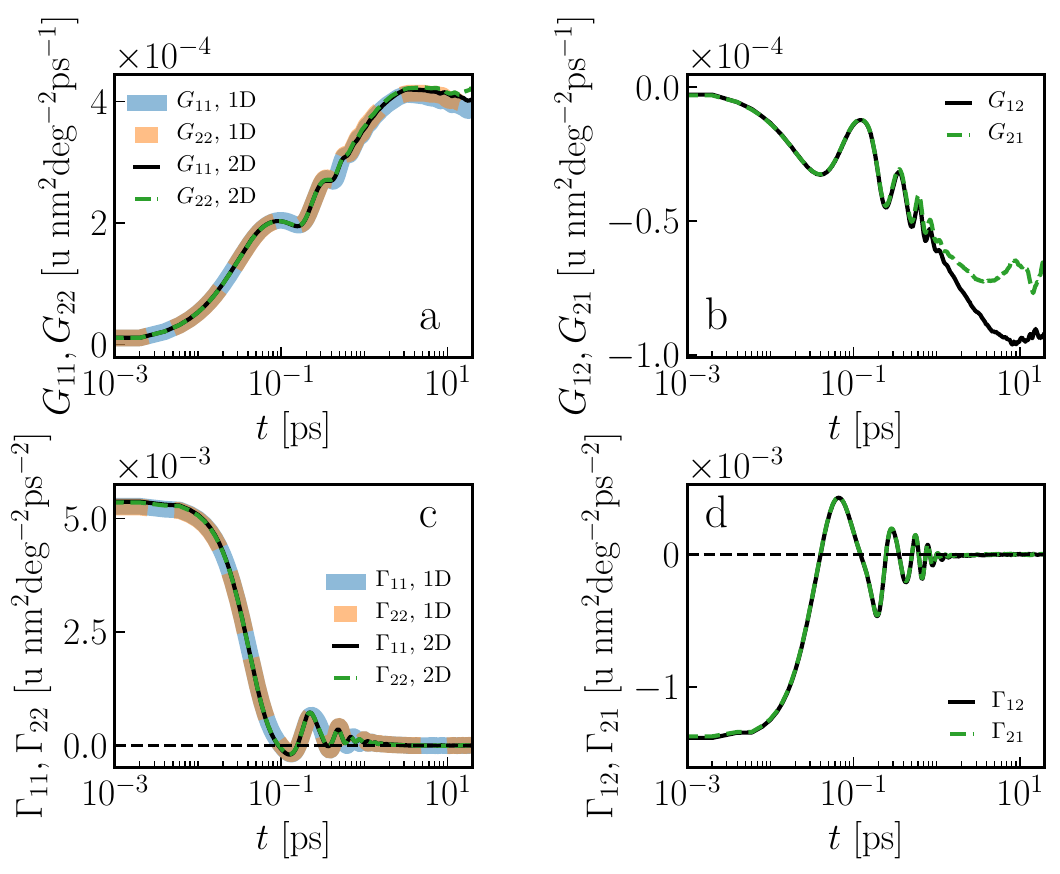}
	\caption{Extracted memory kernel matrix entries $\Gamma_{kl}$ (c, d) and running integral over the memory kernel matrix entries $G_{kl}$ (a, b) from 2D trajectories $\theta_1(t)$ (1) and $\theta_2(t)$ (2) for pentane (compare figure~\ref{fig:matrix_exp_fit_ikernels_2D_dih_pentane}), computed using equation~\eqref{eq:G-iter} (black and green lines). The light blue and orange lines in (a, c) are 1D extraction results (section \ref{sec:mapping_1D}).}
	\label{fig:matrix_exp_fit_ikernels_2D_dih_pentane_all}
\end{figure*}

Figure~\ref{fig:matrix_exp_fit_ikernels_2D_dih_pentane_all} depicts all entries of the extracted memory kernel matrix for pentane (compare figure~\ref{fig:matrix_exp_fit_ikernels_2D_dih_pentane}). As expected (appendix \ref{app:corr_pentane}), the diagonal entries are very similar to each other. The same is true for the off-diagonal entries. We see small deviations for $G_{12}$ and $G_{21}$ in figure~\ref{fig:matrix_exp_fit_ikernels_2D_dih_pentane_all}(b) for long times, presumably due to the finite MD simulation length.

\section{Markovian embedding simulations with exponential-oscillatory memory kernel}
\label{app:exp_osc_1D_me}

\begin{figure*}
	\centering
	\includegraphics[width=1
	\linewidth]{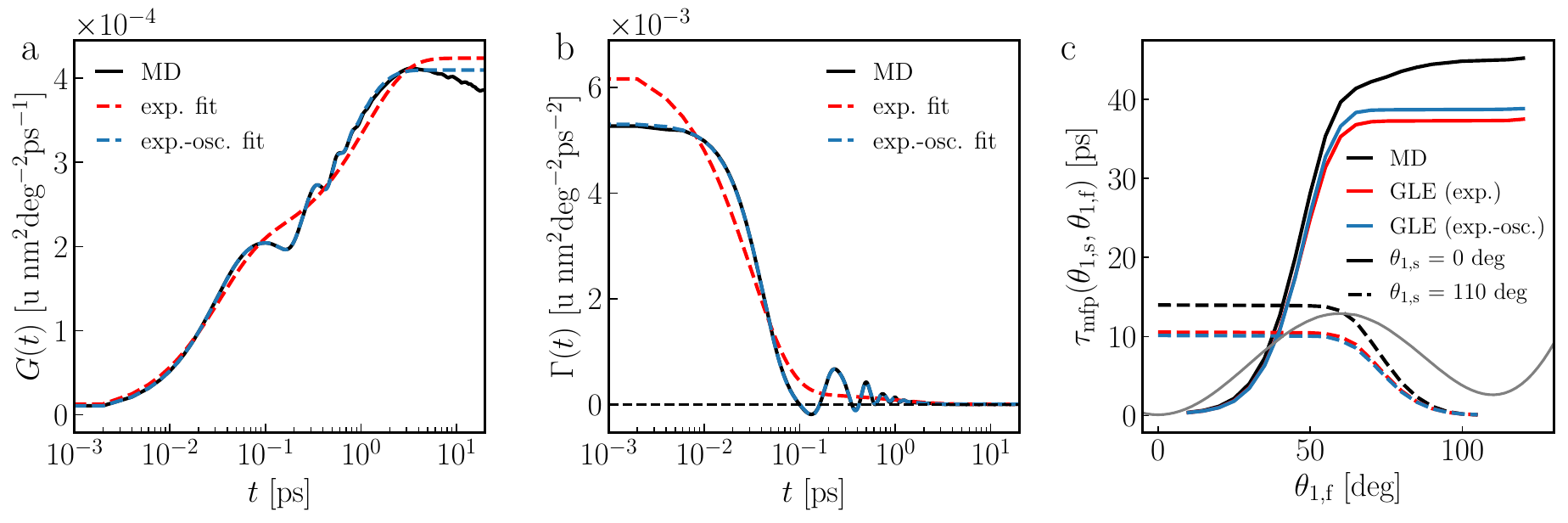}
	\caption{Comparison between different 1D Markovian embedding schemes for the $\theta_1$ dihedral angle of pentane. (a, b) Extracted 1D memory kernel $\Gamma(t)$ (b) and running integral over the memory kernel $G(t)$ (a), also shown in figure~\ref{fig:matrix_exp_fit_ikernels_2D_dih_pentane_all}. The broken lines are fits according to equation~\eqref{eq:fit_kernel} for a sum of $m=5$ exponentials (red)  and according to equation~\eqref{eq:fit_kernel_exposc} for a sum of $m=5$ exponential-oscillatory components (blue) to the data. The fitting parameters are summarized in appendix \ref{app:fitting_constants_1D}. (c) Comparison between the MFPT $\tau_\text{mfp}$ from MD (black) and 1D GLE simulations as a function of the final position $\theta_\text{1,f}$ for different starting positions $\theta_\text{1,s}$, here for the trans-to-cis transition ($\theta_\text{1,s}$ = 0 deg, solid lines) and cis-to-trans transition ($\theta_\text{1,s}$ = 110 deg, broken lines) The 1D GLE results stem from simulations according to equations~\eqref{eq:markov_embedding_1D} (red) and equations~\eqref{eq:markov_embedding_1D_exposc} (blue) with the parameters given in appendix \ref{app:fitting_constants_1D}. The gray curve shows the 1D potential landscape from figure~\ref{fig:pentane_dihedrals_scheme}. }
	\label{fig:ME_1D_pentane_kernel_dihedral1}
\end{figure*}
A perfect fit of the 1D memory kernels $\Gamma(t)$ for pentane dihedral angles using multi-exponential functions (equation~\eqref{eq:fit_kernel}) is not possible due to oscillations in the data (compare figure~\ref{fig:matrix_exp_fit_ikernels_2D_dih_pentane} and  figure~\ref{fig:ME_1D_pentane_kernel_dihedral1}). To account for these features, a sum of $m$ exponential-oscillatory components representing the memory kernel was proposed \cite{brunig2022timedependent, brunig2022pair}
\begin{equation}
	\label{eq:fit_kernel_exposc}
	\Gamma(t) = \sum_{i=1}^{m} k_ie^{-t/\tau_{i}}\left[\frac{1}{\tau_iw_i}\sin{(w_i t)}+ \cos{(w_i t)}\right].
\end{equation}
The GLE for $x = \theta_1$ and $m_x = M_{\theta_1\theta_1}$ in equation~\eqref{eq:1D_GLE} is then equivalent to the following system of Markovian Langevin equations:
\begin{align}
	\label{eq:markov_embedding_1D_exposc}
	m_x\Ddot{x}(t) =& - \nabla U\bigl(x(t)\bigr) - \sum_{i=1}^{m} k_i\bigl(x(t) - y_i(t)\bigr)  , \\ \nonumber
	m_{y}^i \Ddot{y}_i(t) &= - \gamma_i \Dot{y}_i(t) + k_i\bigl(x(t) - y_i(t)\bigr) + \zeta_i(t),
\end{align}
with $U(x) = -k_BT \ln \rho(x)$ and where $m_y^i$ and $\gamma_i$ correspond to the parameters $k_i$, $\tau_i$ and $w_i$ via
\begin{align}
	m_y^i &= \frac{k_i}{\tau_i^{-2} + w_i^2}, \\
	\gamma_i &= 2\frac{m_y^i}{\tau_i}.
\end{align}
As visible in figure~\ref{fig:ME_1D_pentane_kernel_dihedral1}(a, b), the model in equation~\eqref{eq:fit_kernel_exposc} for $m=5$  matches the extracted 1D memory kernel much better as the multi-exponential Ansatz in equation~\eqref{eq:fit_kernel}. The fit parameters of the $m=5$ components are given in appendix \ref{app:fitting_constants_1D}. With the fitting parameters, we numerically solve the Markovian embedding in equations~\eqref{eq:markov_embedding_1D_exposc} with to simulate $x(t) = \theta_1(t)$, with $\zeta_i(t)$ drawn from Gaussian processes with zero mean and $\langle \zeta_i(t) \zeta_j(t')\rangle = 2  k_BT \gamma_i \delta_{ij} \delta(|t-t' |)$. Despite better fitting of the memory kernel, this GLE model does not improve the MFPT predictions, as shown in figure~\ref{fig:ME_1D_pentane_kernel_dihedral1}(c). The trans-to-cis transition is slightly slowed down compared to data in red, but no improvement is observed in the cis-to-trans transition.

\section{Analytical expression of a two-dimensional exponential memory matrix}
\label{app:comp_kernel_matrix}
In the following, we derive the explicit components for the memory kernel matrix $\Hat{\Gamma}_{i}(t)$ in equation~\eqref{eq:memory_matrix} for a two-dimensional coordinate ($n=2$). The entries for a single-exponential component $\Hat{\Gamma}_{i}(t)$ read
\begin{align}
	\Hat{\Gamma}_{i}(t) =& \left( \begin{array}{rr} 
		\gamma_i^{11} & \gamma_i^{12}  \\ 
		\gamma_i^{21} & \gamma_i^{22}  \\ 
	\end{array}\right)  \cdot \left( \begin{array}{rr} 
		\tau_i^{11} & \tau_i^{12}  \\ 
		\tau_i^{21} & \tau_i^{22}  \\ 
	\end{array}\right)^{-1} \cdot e^{-t\left( \begin{array}{rr} 
			\tau_i^{11} & \tau_i^{12}  \\ 
			\tau_i^{21} & \tau_i^{22}\end{array}\right)^{-1}},\\ \nonumber
	=& \frac{1}{A}\left(\begin{array}{rr} 
		\gamma_i^{11}\tau_i^{22}  - \gamma_i^{12}\tau_i^{21}   &  \gamma_i^{12}\tau_i^{11} - \gamma_i^{11}\tau_i^{12}  \\ 
		\gamma_i^{21}\tau_i^{22}  - \gamma_i^{22}\tau_i^{21}   &  \gamma_i^{22}\tau_i^{11} - \gamma_i^{21}\tau_i^{12} 
	\end{array}\right) \\ \nonumber & \cdot e^{-\frac{t}{A}\left( \begin{array}{rr} 
			\tau_i^{22} & -\tau_i^{12}  \\ 
			-\tau_i^{21} & \tau_i^{11}\end{array}\right)},
\end{align}
where $A = \text{det}(\Hat{\tau}_{i})$.
To arrive at the explicit form of the memory kernel entries, we have to diagonalize the matrix exponential $exp(t\hat{C})$, where $\hat{C} = -\Hat{\tau}_{i}^{-1}$. If $\hat{C}$ is diagonalizable with the diagonal matrix $\hat{D}$ and the eigenbasis $\hat{V}$, it follows that
\begin{equation}
	\label{eq:exp_matrix}
	e^{t\hat{C}} = \hat{V} e^{t\hat{D}} \hat{V}^{-1}.
\end{equation}
The eigenvalues following from the characteristic polynomial are $\lambda_{1,2} = -\frac{1}{2A}[(\tau_{i}^{11} + \tau_{i}^{22}) \mp \sqrt{(\tau_{i}^{11} - \tau_{i}^{22})^2 + 4(\tau_{i}^{21}\tau_{i}^{12})}]$. 
For the two eigenvectors or the eigenbasis, we have
\begin{equation}
	\hat{V} = \frac{\sqrt{2}}{2}\left(\begin{array}{rr} 
		1& 1  \\ 
		-1 & 1  \\ 
	\end{array}\right). 
\end{equation}
Note, that we normalized $\hat{V}$ such that the sum of the squares of the elements of each eigenvector equals unity.
Using the found eigenvalues and eigenbasis of the matrix $\hat{C}$ and equation~\eqref{eq:exp_matrix}, we obtain the analytical form for the memory kernel matrix entries in equation~\eqref{eq:memory_matrix}
\begin{align}
	\label{eq:memory_matrix_model2}
	\Hat{\Gamma}_{i}(t) =& \frac{1}{2A} \left( \begin{array}{rr} 
		\gamma_i^{11}\tau_i^{22}  - \gamma_i^{12}\tau_i^{21}   &  \gamma_i^{12}\tau_i^{11} - \gamma_i^{11}\tau_i^{12}  \\ 
		\gamma_i^{21}\tau_i^{22}  - \gamma_i^{22}\tau_i^{21}   &  \gamma_i^{22}\tau_i^{11} - \gamma_i^{21}\tau_i^{12} 
	\end{array}\right) \\ \nonumber & \cdot \left( \begin{array}{rr} 
		e^{t\lambda_1} +  e^{t\lambda_2} & e^{t\lambda_2} -  e^{t\lambda_1} \\ 
		e^{t\lambda_2} -  e^{t\lambda_1}  & e^{t\lambda_1} +  e^{t\lambda_2} 
	\end{array}\right).
\end{align}
From equation~\eqref{eq:memory_matrix} we recall that the friction coefficient matrix gives the long-time limits of the $\hat{G}(t)$ entries, i.e. $\Hat{G}_i(t\rightarrow \infty) = \int_0^{\infty} ds\:\Hat{\Gamma}_i(s) = \Hat{\gamma}_i$. All entries of $\Hat{\Gamma}_i$ in equation~\eqref{eq:memory_matrix_model2} consist of sums of exponential functions depending on diagonal and off-diagonal memory times. For low off-diagonal memory times, i.e.  $\tau_{12}, \tau_{21} \rightarrow 0$, and low off-diagonal friction, i.e. $\gamma_{12}, \gamma_{21} \rightarrow 0$, the diagonal entries in equation~\eqref{eq:memory_matrix_model2} reduce to uncoupled single-exponentially decaying functions (compare equation~\eqref{eq:fit_kernel}), and the off-diagonal entries vanish. For $\tau_{12}, \tau_{21} \rightarrow 0$, but $\gamma_{12}, \gamma_{21} \neq 0$, we find
\begin{align}
	\Hat{\Gamma}_{i}(t) = \left( \begin{array}{rr} 
		\frac{\gamma_i^{11}}{\tau_i^{11}} e^{-t/\tau_i^{11}}  &  \frac{\gamma_i^{12}}{\tau_i^{22}} e^{-t/\tau_i^{22}}  \\ 
		\frac{\gamma_i^{21}}{\tau_i^{11}} e^{-t/\tau_i^{11}}  &  \frac{\gamma_i^{22}}{\tau_i^{22}} e^{-t/\tau_i^{22}}
	\end{array}\right).
\end{align}
For vanishing off-diagonal memory times, the diagonal entries in the memory kernel matrix reduce to uncoupled 1D memory kernels as given in equation~\eqref{eq:fit_kernel}. Off-diagonal entries decay with the diagonal memory times. 
For pentane (figure~\ref{fig:matrix_exp_fit_ikernels_2D_dih_pentane}) and alanine dipeptide (figure~\ref{fig:matrix_exp_fit_kernels_2D_dih_ala2}), the diagonal entries of the memory matrix, in fact, match 1D extraction results. Also, the fitted off-diagonal memory times from the memory kernel data shown in appendix \ref{app:fitting_constants_2D} are small, which is consistent with the findings in this appendix.
\section{Fitting constants of the one-dimensional memory kernel fits}
\label{app:fitting_constants_1D}
This appendix summarizes the 1D memory kernel fitting parameters in tables 
\ref{tab:fit_kernel_1D_pentane} -~\ref{tab:fit_kernel_1D_ala2}.
\begin{table}[!htbp]
	\centering
	\caption{Fitting parameters for the 1D memory kernel fit according to equation~\eqref{eq:fit_kernel} for the pentane MD data shown in figure~\ref{fig:matrix_exp_fit_ikernels_2D_dih_pentane_all}; for a representation of the fit for $\theta_1$, see figure~\ref{fig:ME_1D_pentane_kernel_dihedral1}.}
	\begin{ruledtabular}
		\begin{tabular}{c  c  c  c  c} 
			
			Parameter & $\theta_1$ & $\theta_2$ \\ 
			\hline\\[-2.3ex]
			$\gamma_{1}$  [10$^{-4}$ u nm$^{2}$ deg$^{-2}$ ps$^{-1}$] &  0.13 & 0.72 \\
			$\gamma_{2}$  [10$^{-4}$ u nm$^{2}$ deg$^{-2}$ ps$^{-1}$] &  0.05  & 0.61  \\
			$\gamma_{3}$ [10$^{-4}$ u nm$^{2}$ deg$^{-2}$ ps$^{-1}$] &  1.99 & 1.70\\
			$\gamma_{4}$  [10$^{-4}$ u nm$^{2}$ deg$^{-2}$ ps$^{-1}$] &  0.05 & 0.53 \\
			$\gamma_{5}$  [10$^{-4}$ u nm$^{2}$ deg$^{-2}$ ps$^{-1}$] &  2.03 & 0.60 \\
			\hline
			$\tau_{1}$ [ps] & 1.11 & 0.73  \\
			$\tau_{2}$ [ps] & 1.11  & 0.73 \\
			$\tau_{3}$ [ps]&  0.03  & 0.02 \\
			$\tau_{4}$ [ps] &  1.11 & 0.73 \\
			$\tau_{5}$ [ps]&  1.11  & 0.73 \\
		\end{tabular}
	\end{ruledtabular}
	\label{tab:fit_kernel_1D_pentane}
\end{table}

\begin{table}[!htbp]
	\centering
	\caption{Fitting parameters for the 1D memory kernel fit according to equation~\eqref{eq:fit_kernel_exposc} for the pentane dihedral angle ($\theta_1$) shown in figure~\ref{fig:ME_1D_pentane_kernel_dihedral1} in appendix \ref{app:exp_osc_1D_me}.}
	\begin{ruledtabular}
		\begin{tabular}{c  c  c  c } 
			
			$i$ & $k_i$ [10$^{-3}$ u nm$^{2}$ deg$^{-2}$ ps$^{-2}$] & $\tau_i$ [ps] & $w_i$ [ps$^{-1}$] \\ 
		\hline
			1   &  0.24 & 0.47 & 0.01 \\
			2 &  1.36  & 0.06 & 56.77 \\
			3 &  0.52 & 0.18 & 37.09\\
			4 &  1.34 & 0.27 & 25.31 \\
			5 &  1.86 & 0.05 & 9.35 \\
		\end{tabular}
	\end{ruledtabular}
	\label{tab:fit_kernel_1D_pentane_exposc}
\end{table}

\begin{table}[!htbp]
	\centering
	\caption{Fitting parameters for the 1D memory kernel fit according to equation~\eqref{eq:fit_kernel} for the pentane average ($\theta_\text{avg} = (\theta_1+\theta_2)/2$) and difference ($\theta_\text{diff} = (\theta_1 - \theta_2)/2$) dihedral angle, whose analysis is shown figure~\ref{fig:avg_pentane} in appendix \ref{app:avg_pentane}.}
	\begin{ruledtabular}
		\begin{tabular}{c  c  c  c  c} 
			
			Parameter & $\theta_\text{avg}$ & $\theta_\text{diff}$ \\ 
			\hline\\[-2.3ex]
			$\gamma_{1}$  [10$^{-4}$ u nm$^{2}$ deg$^{-2}$ ps$^{-1}$] &  0.05 & 0.05 \\
			$\gamma_{2}$  [10$^{-4}$ u nm$^{2}$ deg$^{-2}$ ps$^{-1}$] &  17.77  & 23.45  \\
			$\gamma_{3}$ [10$^{-4}$ u nm$^{2}$ deg$^{-2}$ ps$^{-1}$] &  5.28 & 6.54\\
			$\gamma_{4}$  [10$^{-4}$ u nm$^{2}$ deg$^{-2}$ ps$^{-1}$] &  0.05 & 0.05 \\
			$\gamma_{5}$  [10$^{-4}$ u nm$^{2}$ deg$^{-2}$ ps$^{-1}$] &  7.5 & 0.05 \\
			\hline
			$\tau_{1}$ [ps] & 2.02 & 0.79  \\
			$\tau_{2}$ [ps] & 0.69  & 0.76 \\
			$\tau_{3}$ [ps]&  0.04  & 0.03 \\
			$\tau_{4}$ [ps] &  0.9 & 0.77 \\
			$\tau_{5}$ [ps]&  0.69  & 0.78 \\
		\end{tabular}
	\end{ruledtabular}
	\label{tab:fit_kernel_1D_pentane_avg}
\end{table}

\begin{table}[!htbp]
	\centering
	\caption{Fitting parameters for the 1D memory kernel fit according to equation~\eqref{eq:fit_kernel} for the alanine dipeptide data shown in figure~\ref{fig:matrix_exp_fit_kernels_2D_dih_ala2}.}
	\begin{ruledtabular}
		\begin{tabular}{c  c  c  c  c} 
			
			Parameter & $\phi$ & $\psi$ \\ 
		\hline\\[-2.3ex]
			$\gamma_{1}$  [10$^{-3}$ u nm$^{2}$ deg$^{-2}$ ps$^{-1}$] &  0.01 & 0.01  \\
			$\gamma_{2}$  [10$^{-3}$ u nm$^{2}$ deg$^{-2}$ ps$^{-1}$] &  2.68  & 0.01  \\
			$\gamma_{3}$ [10$^{-3}$ u nm$^{2}$ deg$^{-2}$ ps$^{-1}$] &  1.68 & 2.22 \\
			$\gamma_{4}$  [10$^{-3}$ u nm$^{2}$ deg$^{-2}$ ps$^{-1}$] &  0.01 & 2.66  \\
			$\gamma_{5}$  [10$^{-3}$ u nm$^{2}$ deg$^{-2}$ ps$^{-1}$] &  0.40 & 0.70 \\
			\hline
			$\tau_{1}$ [ps] & 4.00 & 1.70 \\
			$\tau_{2}$ [ps] & 0.97  & 1.09 \\
			$\tau_{3}$ [ps]&  0.01  & 0.01 \\
			$\tau_{4}$ [ps] &  0.98 & 1.01 \\
			$\tau_{5}$ [ps]&  4.49  & 1.01 \\
		\end{tabular}
	\end{ruledtabular}
	\label{tab:fit_kernel_1D_ala2}
\end{table}

\section{Transition-path time distributions of pentane}
\label{app:tpt_profiles}
\begin{figure*} 
	\centering
	\includegraphics[width=0.8
	\linewidth]{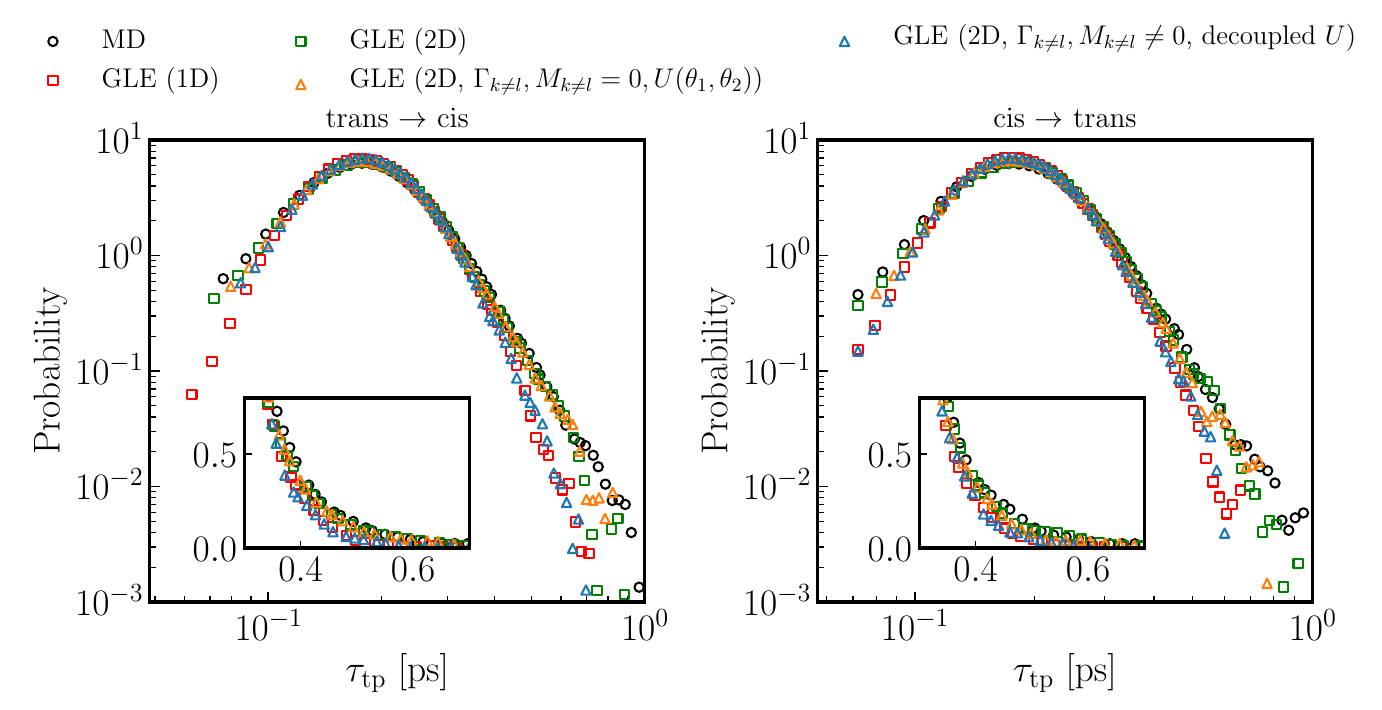}
	\caption{Transition-path time  ($\tau_\text{tp}$) distributions of the pentane dihedral angles ($\theta_1$ and $\theta_2$ averaged) considered in the main text. We show results going from the trans-state (0 deg) to the cis-state (110 deg), and vice versa. We compare distributions computed from the MD simulation (black) with the GLE simulations at different levels of approximation (see figure~\ref{fig:mfpts_profiles_matrix_gle_sim_pentane} for details). The insets show linear plots of the data.}
	\label{fig:pentane_tpt_profiles_2D_gle_sim_dihedrals_mass01_all}
\end{figure*}
Accurately representing non-Markovian systems with Langevin equations requires a precise description of the first-passage time and the transition-path time distribution \cite{dalton2024role}. In figure~\ref{fig:pentane_tpt_profiles_2D_gle_sim_dihedrals_mass01_all}, we present the transition-path time distributions $\tau_\text{tp}$ for the trans- and cis-states of the pentane dihedral angle $\theta_1$, comparing results between MD and GLE simulations (discussed in figure~\ref{fig:mfpts_profiles_matrix_gle_sim_pentane}). A transition path, defined as the trajectory that moves from $x_\text{s}$ to $x_\text{f}$ for the first time without recrossing either state, is considered here.\\
\indent All simulations capture the maximum in both distributions, but only the GLE simulations with the 2D potential (green and orange) accurately predict the tails for short and long transition times. These findings reinforce the conclusions in the main text, highlighting that the 2D potential influences the transition-path dynamics in pentane.
\begin{figure*}
	\centering
	\includegraphics[width=1
	\linewidth]{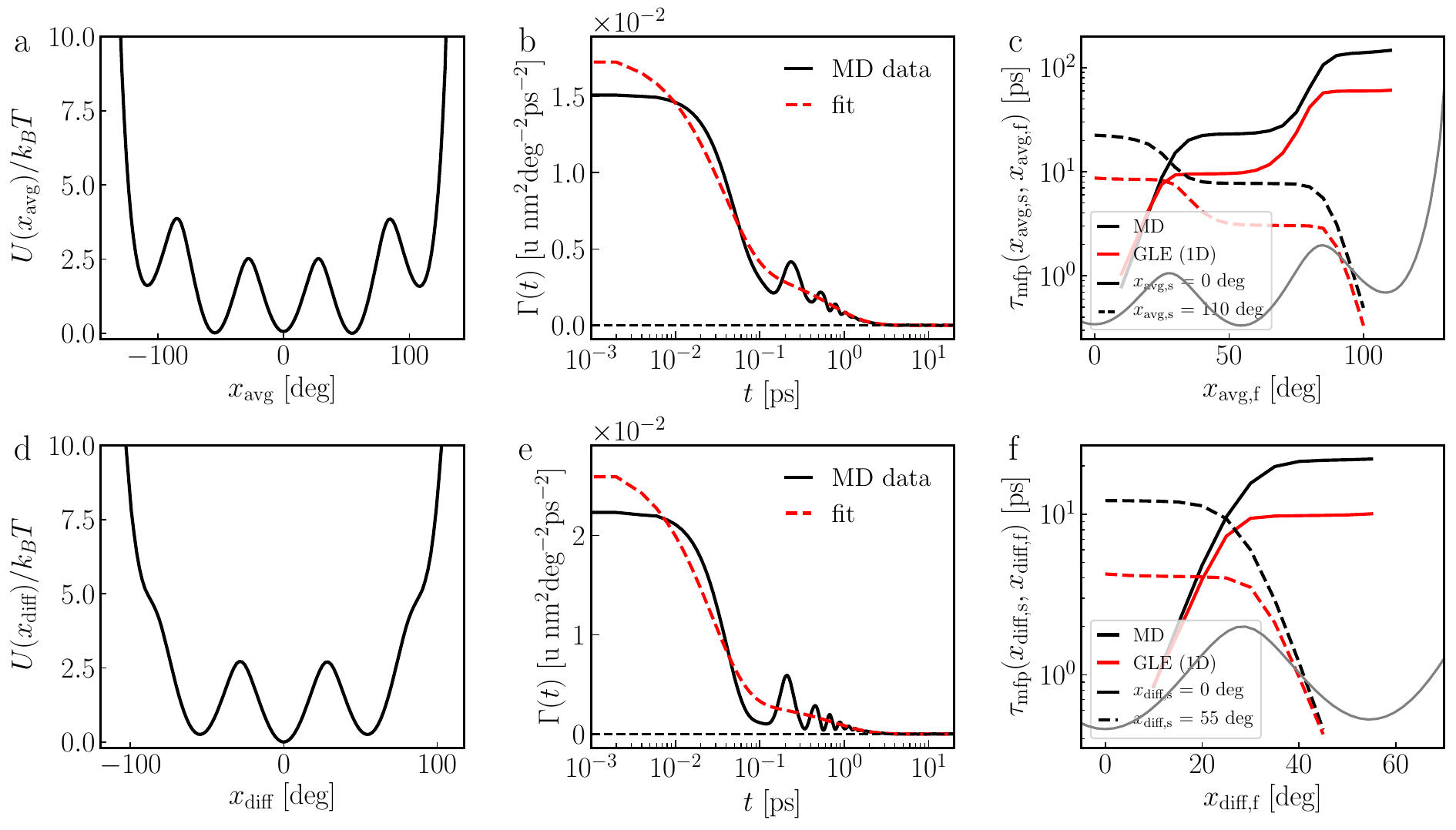}
	\caption{Reduction of the 2D dihedral system of pentane to two alternative 1D reaction coordinates. (a) PMF $U(x_\text{avg}) = - k_BT \ln \rho(x_\text{avg})$ computed using the reaction coordinate $x_\text{avg} = (\theta_1+\theta_2)/2$. (b) Extracted 1D memory kernel $\Gamma(t)$ of the reaction coordinate $x_\text{avg}$ using equation~\eqref{eq:1D_GLE}. The red broken line is a fit according to equation~\eqref{eq:fit_kernel} for a sum of $m=5$ single-exponential components. The fitting parameters are given in appendix \ref{app:fitting_constants_1D}. (c) Comparison between the MFPT $\tau_\text{mfp}$ for the reaction coordinate from MD (black) and 1D GLE simulations (red) as a function of the final position $x_\text{avg,f}$ for different starting positions $x_\text{avg,s}$, here for the trans-to-cis transition ($x_\text{avg,s}$ = 0 deg, solid lines) and cis-to-trans transition ($x_\text{avg,s}$ = 110 deg, broken lines). The 1D GLE results are computed from simulations according to equations~\eqref{eq:markov_embedding_1D} with parameters given in appendix \ref{app:fitting_constants_1D}. The gray curve shows the 1D potential landscape in (a). (d - f) Same analysis for the difference reaction coordinate, i.e. $x_\text{diff} = (\theta_1-\theta_2)/2$. For the computed mean first-passage profiles, we choose the starting points at $x_\text{diff,s}$ = 0 deg (solid lines) and at $x_\text{diff,s}$ = 55 deg (broken lines).}
	\label{fig:avg_pentane}
\end{figure*}

\begin{figure*} 
	\centering
	\includegraphics[width=0.66
	\linewidth]{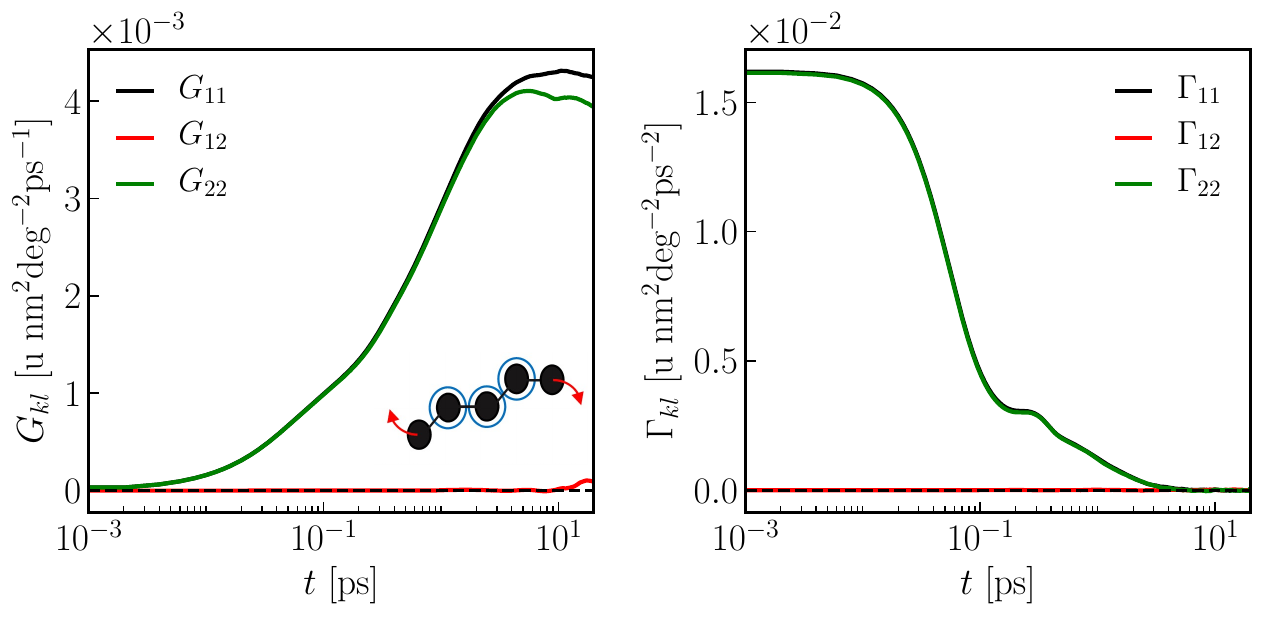}
	\caption{Extracted memory kernel matrix entries $\Gamma_{kl}$ of the two-dimensional dihedral angles trajectory of pentane, $\theta_1(t)$ (1) and $\theta_2(t)$ (2), computed with equation~\eqref{eq:G-iter}. We show results for an MD simulation of pentane where the three inner carbon atoms are frozen.}
	\label{fig:kernels_2D_dih_pentane_m_visc_3constr}
\end{figure*}  

\section{Reduction of the pentane dihedral system to alternative 1D coordinates}
\label{app:avg_pentane}

Here, we project the pentane system onto a 1D reaction coordinate. Instead of focusing on just one dihedral angle, we compute the average of both dihedral angles, i.e. $x_\text{avg} \equiv (\theta_1+\theta_2)/2$, to describe the system's transition dynamics using the 1D GLE in equation~\eqref{eq:1D_GLE}. \\
\indent In figure~\ref{fig:avg_pentane}(a), the resulting PMF, $U(x_\text{avg}) = - k_BT \ln \rho(x_\text{avg})$, consists of multiple minima, which are projections of the various trans- and cis-states from the 2D potential shown in figure~\ref{fig:pentane_dihedrals_scheme}. The extracted 1D memory kernel in figure~\ref{fig:avg_pentane}(b) differs significantly from the individual dihedral angle memory kernels in figure~\ref{fig:matrix_exp_fit_ikernels_2D_dih_pentane}.\\
\indent We simulate the 1D GLE via equations~\eqref{eq:markov_embedding_1D} with a memory kernel fit according to equation~\eqref{eq:fit_kernel}. The resulting MFPTs from the 1D GLE simulation in figure~\ref{fig:avg_pentane}(c) show a substantial discrepancy with the MD data. While the two-dimensional potential information appears to be captured in this 1D coordinate (compare figure~\ref{fig:avg_pentane}(a)), the 1D GLE fails to properly model the dihedral dynamics. Presumably, this reaction coordinate introduces strong non-Gaussianity into the random force \cite{mazur1991and,ayaz2022generalized,kiefer2025ngf}.\\
\indent We repeat the analysis for a coordinate defined as the difference between the two dihedral angles, i.e. $x_\text{diff} \equiv (\theta_1-\theta_2)/2$, which is summarized in figure~\ref{fig:avg_pentane}(d - f). The PMF $U(x_\text{diff})$ resembles the individual dihedral potentials, with the cis-states around $\approx \pm\:55$ deg appearing closer to the trans-state (compare figure~\ref{fig:pentane_dihedrals_scheme}). The difference coordinate is also unsuitable for describing the transition dynamics, as demonstrated in figure~\ref{fig:avg_pentane}(f). Thus, only the two-dimensional dihedral system presented in the main text yields an accurate prediction of the MFPTs for both dihedral angles.

\section{Results for pentane with 3 fixed carbon atoms}
\label{sec:pentane_3constr}
In figure~\ref{fig:kernels_2D_dih_pentane_m_visc_3constr}, we present the extracted memory kernel matrix $\hat{\Gamma}(t)$ for pentane from MD simulations (see appendix \ref{app:simulation_setup}), where the three inner carbon atoms are fixed in space and only the outer carbon atoms are allowed to rotate. Similar to the freely moving scenario shown in figure~\ref{fig:matrix_exp_fit_ikernels_2D_dih_pentane}, the diagonal entries are identical, and, as previously noted for a butane molecule \cite{daldrop2018butane}, the oscillations in the constrained case are suppressed relative to the freely moving scenario. Importantly, the off-diagonal entry $\Gamma_{12}(t)$ is negligible, in contrast to the freely moving case shown in figure~\ref{fig:matrix_exp_fit_ikernels_2D_dih_pentane}. In this constrained scenario, internal friction is diminished \cite{daldrop2018butane}, suggesting a correlation between internal friction effects, due to the motion of the three inner carbon atoms, and off-diagonal friction coupling.

\newpage
\bibliographystyle{unsrt}
\bibliography{sources}

\begin{thebibliography}{10}

\bibitem{van1998remarks}
N.~G. van Kampen.
\newblock {Remarks on Non-Markov Processes}.
\newblock {\em Brazilian Journal of Physics}, 28:90--96, 1998.

\bibitem{huang2009introduction}
K.~Huang.
\newblock {\em {Introduction to Statistical Physics}}.
\newblock Chapman and Hall/CRC, 2009.

\bibitem{kardar2007statistical}
M.~Kardar.
\newblock {\em {Statistical Physics of Particles}}.
\newblock Cambridge University Press, 2007.

\bibitem{balian2007microphysics}
R.~Balian.
\newblock {\em {From Microphysics to Macrophysics: Methods and Applications of
  Statistical Physics. Volume II}}.
\newblock Springer Science \& Business Media, 2007.

\bibitem{hijon2010mori}
C.~Hij{\'o}n, P.~Espa{\~n}ol, E.~Vanden-Eijnden, and R.~Delgado-Buscalioni.
\newblock {Mori-Zwanzig Formalism as a Practical Computational Tool}.
\newblock {\em Faraday Discussions}, 144:301--322, 2010.

\bibitem{karimi2012good}
H.~A. Karimi-Varzaneh, N.~F.~A. van Der~Vegt, F.~M{\"u}ller-Plathe, and
  P.~Carbone.
\newblock {How Good are Coarse-Grained Polymer Models? A Comparison for Atactic
  Polystyrene}.
\newblock {\em ChemPhysChem}, 13(15):3428--3439, 2012.

\bibitem{adelman1980generalized}
S.~A. Adelman.
\newblock {Generalized Langevin Theory for Many-Body Problems in Chemical
  Dynamics: Reactions in Liquids}.
\newblock {\em The Journal of Chemical Physics}, 73(7):3145--3158, 1980.

\bibitem{straub1987calculation}
J.~E. Straub, M.~Borkovec, and B.~J. Berne.
\newblock {Calculation of Dynamic Friction on Intramolecular Degrees of
  Freedom}.
\newblock {\em Journal of Physical Chemistry}, 91(19):4995--4998, 1987.

\bibitem{canales1998generalized}
M.~Canales and G.~Sese.
\newblock {Generalized Langevin Dynamics Simulations of NaCl Electrolyte
  Solutions}.
\newblock {\em The Journal of Chemical Physics}, 109(14):6004--6011, 1998.

\bibitem{bagchi1983effect}
B.~Bagchi and D.~W. Oxtoby.
\newblock {The Effect of Frequency Dependent Friction on Isomerization Dynamics
  in Solution}.
\newblock {\em The Journal of Chemical Physics}, 78(5):2735--2741, 1983.

\bibitem{plotkin1998non}
S.~S. Plotkin and P.~G. Wolynes.
\newblock {Non-Markovian Configurational Diffusion and Reaction Coordinates for
  Protein Folding}.
\newblock {\em Physical Review Letters}, 80(22):5015, 1998.

\bibitem{satija2019generalized}
R.~Satija and D.~E. Makarov.
\newblock {Generalized Langevin Equation as a Model for Barrier Crossing
  Dynamics in Biomolecular Folding}.
\newblock {\em The Journal of Physical Chemistry B}, 123(4):802--810, 2019.

\bibitem{lange2006collective}
O.~F. Lange and H.~Grubm{\"u}ller.
\newblock {Collective Langevin Dynamics of Conformational Motions in Proteins}.
\newblock {\em The Journal of Chemical Physics}, 124(21):214903, 2006.

\bibitem{zwanzig1961memory}
R.~Zwanzig.
\newblock {Memory Effects in Irreversible Thermodynamics}.
\newblock {\em Physical Review}, 124(4):983, 1961.

\bibitem{mori1965transport}
H.~Mori.
\newblock {Transport, Collective Motion, and Brownian Motion}.
\newblock {\em Progress of Theoretical Physics}, 33(3):423--455, 1965.

\bibitem{carof2014two}
A.~Carof, R.~Vuilleumier, and B.~Rotenberg.
\newblock {Two Algorithms to Compute Projected Correlation Functions in
  Molecular Dynamics Simulations}.
\newblock {\em The Journal of Chemical Physics}, 140(12):124103, 2014.

\bibitem{chorin2002optimal}
A.~J. Chorin, O.~H. Hald, and R.~Kupferman.
\newblock {Optimal Prediction with Memory}.
\newblock {\em Physica D: Nonlinear Phenomena}, 166(3-4):239--257, 2002.

\bibitem{ayaz2021non}
C.~Ayaz, L.~Tepper, F.~N. Br{\"u}nig, J.~Kappler, J.~O. Daldrop, and R.~R.
  Netz.
\newblock {Non-Markovian Modeling of Protein Folding}.
\newblock {\em Proceedings of the National Academy of Sciences},
  118(31):e2023856118, 2021.

\bibitem{klippenstein2021introducing}
V.~Klippenstein, M.~Tripathy, G.~Jung, F.~Schmid, and N.~F.~A. van~der Vegt.
\newblock {Introducing Memory in Coarse-Grained Molecular Simulations}.
\newblock {\em The Journal of Physical Chemistry B}, 125(19):4931--4954, 2021.

\bibitem{brunig2022timedependent}
F.~N. Br{\"u}nig, O.~Geburtig, A.~von Canal, J.~Kappler, and R.~R. Netz.
\newblock {Time-Dependent Friction Effects on Vibrational Infrared Frequencies
  and Line Shapes of Liquid Water}.
\newblock {\em The Journal of Physical Chemistry B}, 126(7):1579--1589, 2022.

\bibitem{brunig2022pair}
F.~N. Br{\"u}nig, J.~O. Daldrop, and R.~R. Netz.
\newblock {Pair-Reaction Dynamics in Water: Competition of Memory, Potential
  Shape, and Inertial Effects}.
\newblock {\em The Journal of Physical Chemistry B}, 126(49):10295--10304,
  2022.

\bibitem{dalton2022protein}
B.~A. Dalton, C.~Ayaz, H.~Kiefer, A.~Klimek, L.~Tepper, and R.~R. Netz.
\newblock {Fast Protein Folding is Governed by Memory-Dependent Friction}.
\newblock {\em Proceedings of the National Academy of Sciences},
  120(31):e2220068120, 2023.

\bibitem{dalton2024role}
B.~A. Dalton, H.~Kiefer, and R.~R. Netz.
\newblock {The Role of Memory-Dependent Friction and Solvent Viscosity in
  Isomerization Kinetics in Viscogenic Media}.
\newblock {\em Nature Communications}, 15(1):3761, 2024.

\bibitem{dalton2024memory}
B.~A. Dalton, A.~Klimek, H.~Kiefer, F.~N. Br{\"u}nig, H.~Colinet, L.~Tepper,
  A.~Abbasi, and R.~R. Netz.
\newblock {Memory and Friction: From the Nanoscale to the Macroscale}.
\newblock {\em Annual Review of Physical Chemistry}, 76:431--454, 2025.

\bibitem{kiefer2024predictability}
H.~Kiefer, D.~Furtel, C.~Ayaz, A.~Klimek, J.~O. Daldrop, and R.~R. Netz.
\newblock {Predictability Analysis and Prediction of Discrete Weather and
  Financial Time-Series Data with a Hamiltonian-Based Filter-Projection
  Approach}.
\newblock {\em arXiv preprint arXiv:2409.15026}, 2024.

\bibitem{berne1990dynamic}
B.~J. Berne, M.~E. Tuckerman, J.~E. Straub, and A.~L.~R. Bug.
\newblock {Dynamic Friction on Rigid and Flexible Bonds}.
\newblock {\em The Journal of Chemical Physics}, 93(7):5084--5095, 1990.

\bibitem{ceriotti2010colored}
M.~Ceriotti, G.~Bussi, and M.~Parrinello.
\newblock {Colored-Noise Thermostats {\`a} la Carte}.
\newblock {\em Journal of Chemical Theory and Computation}, 6(4):1170--1180,
  2010.

\bibitem{ayaz2022generalized}
C.~Ayaz, L.~Scalfi, B.~A. Dalton, and R.~R. Netz.
\newblock {Generalized Langevin Equation with a Nonlinear Potential of Mean
  Force and Nonlinear Memory Friction From a Hybrid Projection Scheme}.
\newblock {\em Physical Review E}, 105(5):054138, 2022.

\bibitem{vroylandt2022gle}
H.~Vroylandt.
\newblock {On the Derivation of the Generalized Langevin Equation and the
  Fluctuation-Dissipation Theorem}.
\newblock {\em Europhysics Letters}, 140(6):62003, 2022.

\bibitem{best2013native}
R.~B. Best, G.~Hummer, and W.~A. Eaton.
\newblock {Native Contacts Determine Protein Folding Mechanisms in Atomistic
  Simulations}.
\newblock {\em Proceedings of the National Academy of Sciences},
  110(44):17874--17879, 2013.

\bibitem{berezhkovskii2018single}
A.~M. Berezhkovskii and D.~E. Makarov.
\newblock {Single-Molecule Test for Markovianity of the Dynamics Along a
  Reaction Coordinate}.
\newblock {\em The Journal of Physical Chemistry Letters}, 9(9):2190--2195,
  2018.

\bibitem{abrash1990viscosity}
S.~Abrash, S.~Repinec, and R.~M. Hochstrasser.
\newblock {The Viscosity Dependence and Reaction Coordinate for Isomerization
  of Cis-Stilbene}.
\newblock {\em The Journal of Chemical Physics}, 93(2):1041--1053, 1990.

\bibitem{bagchi2012molecular}
B.~Bagchi.
\newblock {\em {Molecular Relaxation in Liquids}}.
\newblock Oxford University Press, 2012.

\bibitem{langer1969statistical}
J.~S. Langer.
\newblock {Statistical Theory of the Decay of Metastable States}.
\newblock {\em Annals of Physics}, 54(2):258--275, 1969.

\bibitem{grote1981reactive}
R.~F. Grote and J.~T. Hynes.
\newblock {Reactive Modes in Condensed Phase Reactions}.
\newblock {\em The Journal of Chemical Physics}, 74(8):4465--4475, 1981.

\bibitem{van1981stochastic}
W.~F. van Gunsteren, H.~J.~C. Berendsen, and J.~A.~C. Rullmann.
\newblock {Stochastic Dynamics for Molecules with Constraints: Brownian
  Dynamics of n-Alkanes}.
\newblock {\em Molecular Physics}, 44(1):69--95, 1981.

\bibitem{acharya2023diffusion}
S.~Acharya and B.~Bagchi.
\newblock {Diffusion in a Two-Dimensional Energy Landscape in the Presence of
  Dynamical Correlations and Validity of Random Walk Model}.
\newblock {\em Physical Review E}, 107(2):024127, 2023.

\bibitem{mazur1991and}
P.~Mazur and D.~Bedeaux.
\newblock {When and Why is the Random Force in Brownian Motion a Gaussian
  Process}.
\newblock {\em Biophysical Chemistry}, 41(1):41--49, 1991.

\bibitem{vroylandt2022position}
H.~Vroylandt and P.~Monmarch{\'e}.
\newblock {Position-Dependent Memory Kernel in Generalized Langevin Equations:
  Theory and Numerical Estimation}.
\newblock {\em The Journal of Chemical Physics}, 156(24):244105, 2022.

\bibitem{van1982reactive}
G.~van~der Zwan and J.~T. Hynes.
\newblock {Reactive Paths in the Diffusion Limit}.
\newblock {\em The Journal of Chemical Physics}, 77(3):1295--1301, 1982.

\bibitem{bryngelson1989intermediates}
J.~D. Bryngelson and P.~G. Wolynes.
\newblock {Intermediates and Barrier Crossing in a Random Energy Model (with
  Applications to Protein Folding)}.
\newblock {\em The Journal of Physical Chemistry}, 93(19):6902--6915, 1989.

\bibitem{langer2000theory}
J.~S. Langer.
\newblock {Theory of the Condensation Point}.
\newblock {\em Annals of Physics}, 281(1-2):941--990, 2000.

\bibitem{kraft2013brownian}
D.~J. Kraft, R.~Wittkowski, B.~ten Hagen, K.~V. Edmond, D.~J. Pine, and
  H.~L{\"o}wen.
\newblock {Brownian Motion and the Hydrodynamic Friction Tensor for Colloidal
  Particles of Complex Shape}.
\newblock {\em Physical Review E}, 88(5):050301, 2013.

\bibitem{acharya2021rate}
S.~Acharya, S.~Mondal, S.~Mukherjee, and B.~Bagchi.
\newblock {Rate of Insulin Dimer Dissociation: Interplay Between Memory Effects
  and Higher Dimensionality}.
\newblock {\em The Journal of Physical Chemistry B}, 125(34):9678--9691, 2021.

\bibitem{acharya2022non}
S.~Acharya and B.~Bagchi.
\newblock {Non-Markovian Rate Theory on a Multidimensional Reaction Surface:
  Complex Interplay Between Enhanced Configuration Space and Memory}.
\newblock {\em The Journal of Chemical Physics}, 156(13):134101, 2022.

\bibitem{deutch1971molecular}
J.~M. Deutch and I.~Oppenheim.
\newblock {Molecular Theory of Brownian Motion for Several Particles}.
\newblock {\em The Journal of Chemical Physics}, 54(8):3547--3555, 1971.

\bibitem{ermak1978brownian}
D.~L. Ermak and J.~A. McCammon.
\newblock {Brownian Dynamics with Hydrodynamic Interactions}.
\newblock {\em The Journal of Chemical Physics}, 69(4):1352--1360, 1978.

\bibitem{reichert2004hydrodynamic}
M.~Reichert and H.~Stark.
\newblock {Hydrodynamic Coupling of Two Rotating Spheres Trapped in Harmonic
  Potentials}.
\newblock {\em Physical Review E}, 69(3):031407, 2004.

\bibitem{jung2017frequency}
G.~Jung and F.~Schmid.
\newblock {Frequency-Dependent Hydrodynamic Interaction Between Two Solid
  Spheres}.
\newblock {\em Physics of Fluids}, 29(12):126101, 2017.

\bibitem{ansari1992role}
A.~Ansari, C.~M. Jones, E.~R. Henry, J.~Hofrichter, and W.~A. Eaton.
\newblock {The Role of Solvent Viscosity in the Dynamics of Protein
  Conformational Changes}.
\newblock {\em Science}, 256(5065):1796--1798, 1992.

\bibitem{de2014molecular}
D.~De~Sancho, A.~Sirur, and R.~B. Best.
\newblock {Molecular Origins of Internal Friction Effects on Protein-Folding
  Rates}.
\newblock {\em Nature Communications}, 5(1):1--10, 2014.

\bibitem{daldrop2018butane}
J.~O. Daldrop, J.~Kappler, F.~N. Br{\"u}nig, and R.~R. Netz.
\newblock {Butane Dihedral Angle Dynamics in Water is Dominated by Internal
  Friction}.
\newblock {\em Proceedings of the National Academy of Sciences},
  115(20):5169--5174, 2018.

\bibitem{Hegger09}
R.~Hegger and G.~Stock.
\newblock {Multidimensional Langevin Modeling of Biomolecular Dynamics}.
\newblock {\em The Journal of Chemical Physics}, 130(3):034106, 2009.

\bibitem{Schaudinnus15}
N.~Schaudinnus, B.~Bastian, R.~Hegger, and G.~Stock.
\newblock {Multidimensional Langevin Modeling of Nonoverdamped Dynamics}.
\newblock {\em Physical Review Letters}, 115(5):050602, 2015.

\bibitem{Lickert20}
B.~Lickert and G.~Stock.
\newblock {Modeling Non-Markovian Data Using Markov State and Langevin Models}.
\newblock {\em The Journal of Chemical Physics}, 153(24):244112, 2020.

\bibitem{zwanzig2001nonequilibrium}
R.~Zwanzig.
\newblock {\em {Nonequilibrium Statistical Mechanics}}.
\newblock Oxford University Press, 2001.

\bibitem{lee2019multi}
H.~S. Lee, S.-H. Ahn, and E.~F. Darve.
\newblock {The Multi-Dimensional Generalized Langevin Equation for
  Conformational Motion of Proteins}.
\newblock {\em The Journal of Chemical Physics}, 150(17):174113, 2019.

\bibitem{she2023data}
Z.~She, P.~Ge, and H.~Lei.
\newblock {Data-Driven Construction of Stochastic Reduced Dynamics Encoded with
  Non-Markovian Features}.
\newblock {\em The Journal of Chemical Physics}, 158(3):034102, 2023.

\bibitem{lyu2023construction}
L.~Lyu and H.~Lei.
\newblock {Construction of Coarse-Grained Molecular Dynamics with Many-Body
  Non-Markovian Memory}.
\newblock {\em Physical Review Letters}, 131(17):177301, 2023.

\bibitem{xie2024coarse}
P.~Xie and W.~E.
\newblock {Coarse-Graining Conformational Dynamics with Multidimensional
  Generalized Langevin Equation: How, When, and Why}.
\newblock {\em Journal of Chemical Theory and Computation}, 20(18):7708--7715,
  2024.

\bibitem{xie2024ab}
P.~Xie, R.~Car, and W.~E.
\newblock {Ab Initio Generalized Langevin Equation}.
\newblock {\em Proceedings of the National Academy of Sciences},
  121(14):e2308668121, 2024.

\bibitem{darve2001calculating}
E.~Darve and A.~Pohorille.
\newblock {Calculating Free Energies Using Average Force}.
\newblock {\em The Journal of Chemical Physics}, 115(20):9169--9183, 2001.

\bibitem{berne1970calculation}
B.~J. Berne and G.~D. Harp.
\newblock {On the Calculation of Time Correlation Functions}.
\newblock {\em Advances in Chemical Physics}, 17:63--227, 1970.

\bibitem{kowalik2019memory}
B.~Kowalik, J.~O. Daldrop, J.~Kappler, J.~C.~F. Schulz, A.~Schlaich, and R.~R.
  Netz.
\newblock {Memory-Kernel Extraction for Different Molecular Solutes in Solvents
  of Varying Viscosity in Confinement}.
\newblock {\em Physical Review E}, 100(1):012126, 2019.

\bibitem{Margolis2019}
B.~W.~L. Margolis and K.~R. Lyons.
\newblock {ndsplines: A Python Library for Tensor-Product B-Splines of
  Arbitrary Dimension}.
\newblock {\em Journal of Open Source Software}, 4(42):1745, 2019.

\bibitem{li2017computing}
Z.~Li, H.~S. Lee, E.~Darve, and G.~E. Karniadakis.
\newblock {Computing the Non-Markovian Coarse-Grained Interactions Derived from
  the Mori-Zwanzig Formalism in Molecular Systems: Application to Polymer
  Melts}.
\newblock {\em The Journal of Chemical Physics}, 146(1):014104, 2017.

\bibitem{kappler2019non}
J.~Kappler, V.~B. Hinrichsen, and R.~R. Netz.
\newblock {Non-Markovian Barrier Crossing with Two-Time-Scale Memory is
  Dominated by the Faster Memory Component}.
\newblock {\em The European Physical Journal E}, 42(9):1--16, 2019.

\bibitem{lavacchi2020barrier}
L.~Lavacchi, J.~Kappler, and R.~R. Netz.
\newblock {Barrier Crossing in the Presence of Multi-Exponential Memory
  Functions with Unequal Friction Amplitudes and Memory Times}.
\newblock {\em Europhysics Letters}, 131(4):40004, 2020.

\bibitem{kiefer2025ngf}
H.~Kiefer, B.~J.~A. H\'ery, L.~Tepper, B.~A. Dalton, C.~Ayaz, and R.~R. Netz.
\newblock {Analysis and Simulation of Generalized Langevin Equations with
  Non-Gaussian Orthogonal Forces}.
\newblock {\em arXiv preprint arXiv:2505.15665}, 2025.

\bibitem{mitterwallner2020negative}
B.~G. Mitterwallner, L.~Lavacchi, and R.~R. Netz.
\newblock {Negative Friction Memory Induces Persistent Motion}.
\newblock {\em The European Physical Journal E}, 43(10):1--11, 2020.

\bibitem{klimek2022optimal}
A.~Klimek and R.~R. Netz.
\newblock {Optimal Non-Markovian Composite Search Algorithms for Spatially
  Correlated Targets}.
\newblock {\em Europhysics Letters}, 139(3):32003, 2022.

\bibitem{kappler2018memory}
J.~Kappler, J.~O. Daldrop, F.~N. Br{\"u}nig, M.~D. Boehle, and R.~R. Netz.
\newblock {Memory-Induced Acceleration and Slowdown of Barrier Crossing}.
\newblock {\em The Journal of Chemical Physics}, 148(1):014903, 2018.

\bibitem{smith1993stochastic}
P.~E. Smith, B.~M. Pettitt, and M.~Karplus.
\newblock {Stochastic Dynamics Simulations of the Alanine Dipeptide Using a
  Solvent-Modified Potential Energy Surface}.
\newblock {\em The Journal of Physical Chemistry}, 97(26):6907--6913, 1993.

\bibitem{hummer2003coarse}
G.~Hummer and I.~G. Kevrekidis.
\newblock {Coarse Molecular Dynamics of a Peptide Fragment: Free Energy,
  Kinetics, and Long-Time Dynamics Computations}.
\newblock {\em The Journal of Chemical Physics}, 118(23):10762--10773, 2003.

\bibitem{prada2009exploring}
D.~Prada-Gracia, J.~G{\'o}mez-Garde{\~n}es, P.~Echenique, and F.~Falo.
\newblock {Exploring the Free Energy Landscape: From Dynamics to Networks and
  Back}.
\newblock {\em PLOS Computational Biology}, 5(6):e1000415, 2009.

\bibitem{stamati2010application}
H.~Stamati, C.~Clementi, and L.~E. Kavraki.
\newblock {Application of Nonlinear Dimensionality Reduction to Characterize
  the Conformational Landscape of Small Peptides}.
\newblock {\em Proteins: Structure, Function, and Bioinformatics},
  78(2):223--235, 2010.

\bibitem{leimkuhler2013robust}
B.~Leimkuhler and C.~Matthews.
\newblock {Robust and Efficient Configurational Molecular Sampling via Langevin
  Dynamics}.
\newblock {\em The Journal of Chemical Physics}, 138(17):174102, 2013.

\bibitem{wu2016self}
X.~Wu, B.~R. Brooks, and E.~Vanden-Eijnden.
\newblock {Self-Guided Langevin Dynamics via Generalized Langevin Equation}.
\newblock {\em Journal of Computational Chemistry}, 37(6):595--601, 2016.

\bibitem{kmiecik2016coarse}
S.~Kmiecik, D.~Gront, M.~Kolinski, L.~Wieteska, A.~E. Dawid, and A.~Kolinski.
\newblock {Coarse-Grained Protein Models and Their Applications}.
\newblock {\em Chemical Reviews}, 116(14):7898--7936, 2016.

\bibitem{mardt2018vampnets}
A.~Mardt, L.~Pasquali, H.~Wu, and F.~No{\'e}.
\newblock {VAMPnets for Deep Learning of Molecular Kinetics}.
\newblock {\em Nature Communications}, 9(1):5, 2018.

\bibitem{ayaz_embedding_nl}
C.~Ayaz, L.~Tepper, and R.~R. Netz.
\newblock {Self-Consistent Markovian Embedding of Generalized Langevin
  Equations with Configuration-Dependent Mass and a Nonlinear Friction Kernel}.
\newblock {\em Turkish Journal of Physics}, 46(6):194--205, 2022.

\bibitem{jung2023dynamic}
B.~Jung and G.~Jung.
\newblock {Dynamic Coarse-Graining of Linear and Non-Linear Systems:
  Mori-Zwanzig Formalism and Beyond}.
\newblock {\em The Journal of Chemical Physics}, 159(8):084110, 2023.

\bibitem{wolf2025cross}
N.~Wolf, V.~Klippenstein, and N.~F.~A. van~der Vegt.
\newblock {Cross-Correlations in the Fluctuation--Dissipation Relation
  Influence Barrier-Crossing Dynamics}.
\newblock {\em The Journal of Chemical Physics}, 162(5):054113, 2025.

\bibitem{durlofsky1987dynamic}
L.~Durlofsky, J.~F. Brady, and G.~Bossis.
\newblock {Dynamic Simulation of Hydrodynamically Interacting Particles}.
\newblock {\em Journal of Fluid Mechanics}, 180:21--49, 1987.

\bibitem{happel2012low}
J.~Happel and H.~Brenner.
\newblock {\em {Low Reynolds Number Hydrodynamics: With Special Applications to
  Particulate Media}}, volume~1.
\newblock Springer Science \& Business Media, 2012.

\bibitem{rosenberg1980isomerization}
R.~O. Rosenberg, B.~J. Berne, and D.~Chandler.
\newblock {Isomerization Dynamics in Liquids by Molecular Dynamics}.
\newblock {\em Chemical Physics Letters}, 75(1):162--168, 1980.

\bibitem{pronk2013gromacs}
S.~Pronk, S.~P{\'a}ll, R.~Schulz, P.~Larsson, P.~Bjelkmar, R.~Apostolov, M.~R.
  Shirts, J.~C. Smith, P.~M. Kasson, D.~van~der Spoel, et~al.
\newblock {GROMACS 4.5: A High-Throughput and Highly Parallel Open Source
  Molecular Simulation Toolkit}.
\newblock {\em Bioinformatics}, 29(7):845--854, 2013.

\bibitem{Oostenbrink_2004}
C.~Oostenbrink, A.~Villa, A.~E. Mark, and W.~F. {van Gunsteren}.
\newblock {A Biomolecular Force Field Based on the Free Enthalpy of Hydration
  and Solvation: The GROMOS Force-Field Parameter Sets 53A5 and 53A6}.
\newblock {\em Journal of Computational Chemistry}, 25(13):1656--1676, 2004.

\bibitem{Ryckaert_1977}
J.-P. Ryckaert, G.~Ciccotti, and H.~J.~C. Berendsen.
\newblock {Numerical Integration of the Cartesian Equations of Motion of a
  System with Constraints: Molecular Dynamics of n-Alkanes}.
\newblock {\em Journal of Computational Physics}, 23(3):327--341, 1977.

\bibitem{berendsen1987missing}
H.~J.~C. Berendsen, J.~R. Grigera, and T.~P. Straatsma.
\newblock {The Missing Term in Effective Pair Potentials}.
\newblock {\em Journal of Physical Chemistry}, 91(24):6269--6271, 1987.

\bibitem{Berendsen_1984}
H.~J.~C. Berendsen, J.~P.~M. Postma, W.~F. van Gunsteren, A.~DiNola, and J.~R.
  Haak.
\newblock {Molecular Dynamics with Coupling to an External Bath}.
\newblock {\em The Journal of Chemical Physics}, 81(8):3684--3690, 1984.

\bibitem{bussi2007canonical}
G.~Bussi, D.~Donadio, and M.~Parrinello.
\newblock {Canonical Sampling Through Velocity Rescaling}.
\newblock {\em The Journal of Chemical Physics}, 126(1):014101, 2007.

\bibitem{Darden_1993}
T.~Darden, D.~York, and L.~Pedersen.
\newblock {Particle Mesh Ewald: An N log(N) Method for Ewald Sums in Large
  Systems}.
\newblock {\em The Journal of Chemical Physics}, 98(12):10089--10092, 1993.

\bibitem{salomon2013overview}
R.~Salomon-Ferrer, D.~A. Case, and R.~C. Walker.
\newblock {An Overview of the Amber Biomolecular Simulation Package}.
\newblock {\em Wiley Interdisciplinary Reviews: Computational Molecular
  Science}, 3(2):198--210, 2013.

\bibitem{jorgensen1983comparison}
W.~L. Jorgensen, J.~Chandrasekhar, J.~D. Madura, R.~W. Impey, and M.~L. Klein.
\newblock {Comparison of Simple Potential Functions for Simulating Liquid
  Water}.
\newblock {\em The Journal of Chemical Physics}, 79(2):926--935, 1983.

\bibitem{hess1997lincs}
B.~Hess, H.~Bekker, H.~J.~C. Berendsen, and J.~G. E.~M. Fraaije.
\newblock {LINCS: A Linear Constraint Solver for Molecular Simulations}.
\newblock {\em Journal of Computational Chemistry}, 18(12):1463--1472, 1997.

\bibitem{de2018tidynamics}
P.~de~Buyl.
\newblock {tidynamics: A Tiny Package to Compute the Dynamics of Stochastic and
  Molecular Simulations}.
\newblock {\em Journal of Open Source Software}, 3(28):877, 2018.

\bibitem{2020SciPy-NMeth}
P.~Virtanen, R.~Gommers, T.~E. Oliphant, M.~Haberland, T.~Reddy, D.~Cournapeau,
  E.~Burovski, P.~Peterson, W.~Weckesser, J.~Bright, et~al.
\newblock {SciPy 1.0: Fundamental Algorithms for Scientific Computing in
  Python}.
\newblock {\em Nature Methods}, 17(3):261--272, 2020.

\end{thebibliography}
\end{document}